\documentclass[pra,twocolumn,showpacs]{revtex4}

\usepackage{amssymb}
\usepackage{amsmath}
\usepackage{epsfig}
\newcommand{\br}{{\bf r}}
\newcommand{\bv}{{\bf v}}
\newcommand{\bq}{{\bf q}}

\newcommand{\ba}{{\bf a}}
\newcommand{\bR}{{\bf R}}
\newcommand{\bQ}{{\bf Q}}
\newcommand{\bM}{{\bf M}}
\newcommand{\tbR}{\tilde{{\bf R}}}
\newcommand{\cL}{{\cal L}}
\newcommand{\cF}{{\cal F}}
\newcommand{\wa}{\widetilde{{\bf a}}}

\newcommand{\eqb}{\begin{equation}}
\newcommand{\eqe}{\end{equation}}
\newcommand{\vare}{\varepsilon }

\begin{document}
\title{Nonlinear tunneling in two-dimensional lattices}
\author{V. A. Brazhnyi$^1$, V. V. Konotop$^{1,2}$, V. Kuzmiak$^3$, and V. S. Shchesnovich$^4$}
\affiliation{$^1$Centro de F\'{\i}sica Te\'{o}rica e Computacional,
Universidade de Lisboa,
 Complexo Interdisciplinar, Avenida Professor Gama Pinto 2, Lisboa
1649-003, Portugal\\
$^2$
Departamento de F\'{\i}sica, Faculdade de Ci\^encias,
Universidade de Lisboa, Campo Grande, Ed. C8, Piso 6, Lisboa
1749-016, Portugal.
\\
$^3$ Institute of Photonics and Electronics, v.v.i., Czech
Academy of Sciences, Chaberska 57, 182 51 Prague 8, Czech Republic
\\
$^4$ Instituto de F\'{\i}sica - Universidade Federal de Alagoas, Macei\'o AL
57072-970, Brazil}
%\date{}
\begin{abstract}
We present thorough analysis of the  nonlinear tunneling of Bose-Einstein
condensates in  static and accelerating two-dimensional lattices within the framework of the mean-field approximation. We deal with nonseparable lattices considering different initial atomic distributions
in the highly symmetric states. For analytical description of the condensate before instabilities are developed, we derive several few-mode
models, analyzing both essentially nonlinear and quasi-linear regimes of tunneling. By direct numerical simulations, we show that two-mode models provide accurate description of the tunneling when either initially two states are populated or tunneling occurs between two stable states. Otherwise a two-mode model may give only useful qualitative hints for
understanding tunneling but does not reproduce many features of the phenomenon. This reflects crucial role of the instabilities developed due to two-body interactions resulting in non-negligible population of the higher bands. This effect becomes even more pronounced in the case of accelerating lattices. In the latter case we show that the direction of the acceleration is a relevant physical parameter which affects the tunneling by changing the atomic rates at different symmetric states and by  changing the numbers of bands involved in the atomic transfer.
\end{abstract}

\maketitle

\section{Introduction}

Tunneling in weakly nonlinear systems with periodically varying coefficients is
intrinsically related to instabilities and, therefore,  from the physical point of
view the phenomenon differs fundamentally from its linear counterpart. This has been
suggested in Ref.~\cite{KKS}, where the two-level model for the Landau-Zener (LZ)
tunneling  was introduced, and supported by the direct numerical simulations of the
two-dimensional (2D) Gross-Pitaevskii (GP) equation~\cite{BKK} where resonant
tunneling between bands was stimulated by the nonlinearity rather than by the
linear force and therefore termed as a {\em nonlinear} tunneling. LZ tunneling has also
been extensively studied in a number of experiments carried out with Bose-Einstein
condensates (BECs) loaded in moving optical lattices (OLs)~\cite{exper1,exper2}, where in
particular, it was found that one can distinguish two situations corresponding to
the instability regime and to the regime of Bloch oscillations. In Ref.~\cite{Wimb} destructive
effect of the nonlinearity in Wannier-Stark problem was investigated numerically.
Very recently several experimental results were reported where  resonant tunneling in tilted lattice was observed~\cite{Arimondo} and the role of asymmetry of the OL on the tunneling was studied~\cite{Weitz}.

The problem of tunneling has been addressed theoretically in two different limiting cases. In the case of a shallow lattice the tunneling was considered within the framework of the
phenomenological models suggested in ~\cite{Landau-Zener} and later on thoroughly
analyzed and improved in~\cite{SHCH}. The model accounting spatial dependence of
the realistic field, but still based on the two-level approach was developed
in~\cite{KKS}. While analytical description of the  crossover among the above
limiting cases is hardly achievable, to the best of authors knowledge there exist no
direct numerical simulations allowing one to verify validity of each of the models
and/or to study the transient regimes, and no generalization of the above theories on 2D
and 3D lattices exists as yet, in contrast to the case of the linear tunneling that
is well studied in multidimensional settings~\cite{Kolovsky,LZ2D}. Such a
generalization is the main goal of the present paper.

The work is organized as follows. In
Sec.~\ref{sec:statement} we outline the statement of the problem, presenting
the analytical models with different types of the lattices. In Sec.~\ref{sec:deep}
we deal with various cases of LZ, inter-band, and intra-band tunneling of matter
waves in a deep OL, reducing the GP equation to different
two-- and three--level models. LZ tunneling
in a shallow lattice is considered in Sec.~\ref{sec:shallow}.
Numerical simulations of different types of tunneling in 2D lattices are
described in Sec.~\ref{numerics}. The results obtained are summarized in Conclusion.

\section{Statement of the problem}
\label{sec:statement}

We start with the dimensionless 2D GP equation
\begin{eqnarray}
\label{GPEa}
 i\partial_t\psi =-\Delta\psi+V(\br,t)\psi+  \sigma |\psi|^2\psi,
\end{eqnarray}
where time $t$ and coordinates ${\bf r}=(x,y)$  are measured in units $2/\omega_z$ and $\ell_z=\sqrt{\hbar/m\omega_z}$, respectively, with $\omega_z$ being the linear harmonic oscillator frequency. $\sigma=$sign$\,a_s$ where $a_s$ is the scattering length of the condensate. The dimensionless wave function $\psi({\bf r})$ is
normalized as follows $N=\int |\psi|^2 d^2{\bf r}$, where $N=4\sqrt{2\pi}a_s N_{at}/\ell_z$ and $N_{at}$ is a real number of atoms.
The OL potential reads
\begin{eqnarray}
    V(\br ,t)=2V_0[\cos^2(x-a_xt^2)+\cos^2(y-a_yt^2)
    \nonumber \\
    +2\varepsilon\cos(x-a_xt^2)\cos(y-a_yt^2)],
    \label{OL_accel}
\end{eqnarray}
where the amplitude $V_0$ is measured in terms of recoil energy and $\varepsilon={\bf e}_1\cdot{\bf e}_2$. OL is assumed to be created by the two pairs of the counter-propagating beams with the polarization vectors ${\bf
e}_{1,2}$, which are accelerating in the directions orthogonal to the polarization,
where ${\bf a}=(a_x,a_y)$ denotes the acceleration (for description of experimental realization of multidimensional lattices see e.g.~\cite{2Dlattice}).

For the next consideration it is convenient to make a substitution
\begin{eqnarray}
    \Psi =e^{i(2V_0t+ ({\bf r}-\ba t^2){\bf a} t+a^2t^2/3)} \psi,
\end{eqnarray}
to introduce new independent variables $\tbR={\bf r}-{\bf a}t^2$ and $T=t$, and to rewrite Eq.~(\ref{GPEa}) in the form
\begin{eqnarray}
\label{GPE}
 i \partial_T\Psi =-\Delta_{\tbR} \Psi+({\bf a}\cdot{\tbR})\Psi
 +V(\tbR)\Psi+\sigma  |\Psi|^2\Psi,
\end{eqnarray}
where the potential is time-independent:
\begin{eqnarray}
\label{potential}
V(\tbR)=V_0\left[\cos(2\tilde{X})+\cos(2\tilde{Y})+4\varepsilon\cos(\tilde{X})\cos(\tilde{Y})\right]\,.
\end{eqnarray}

\begin{figure}[h]
\epsfig{file=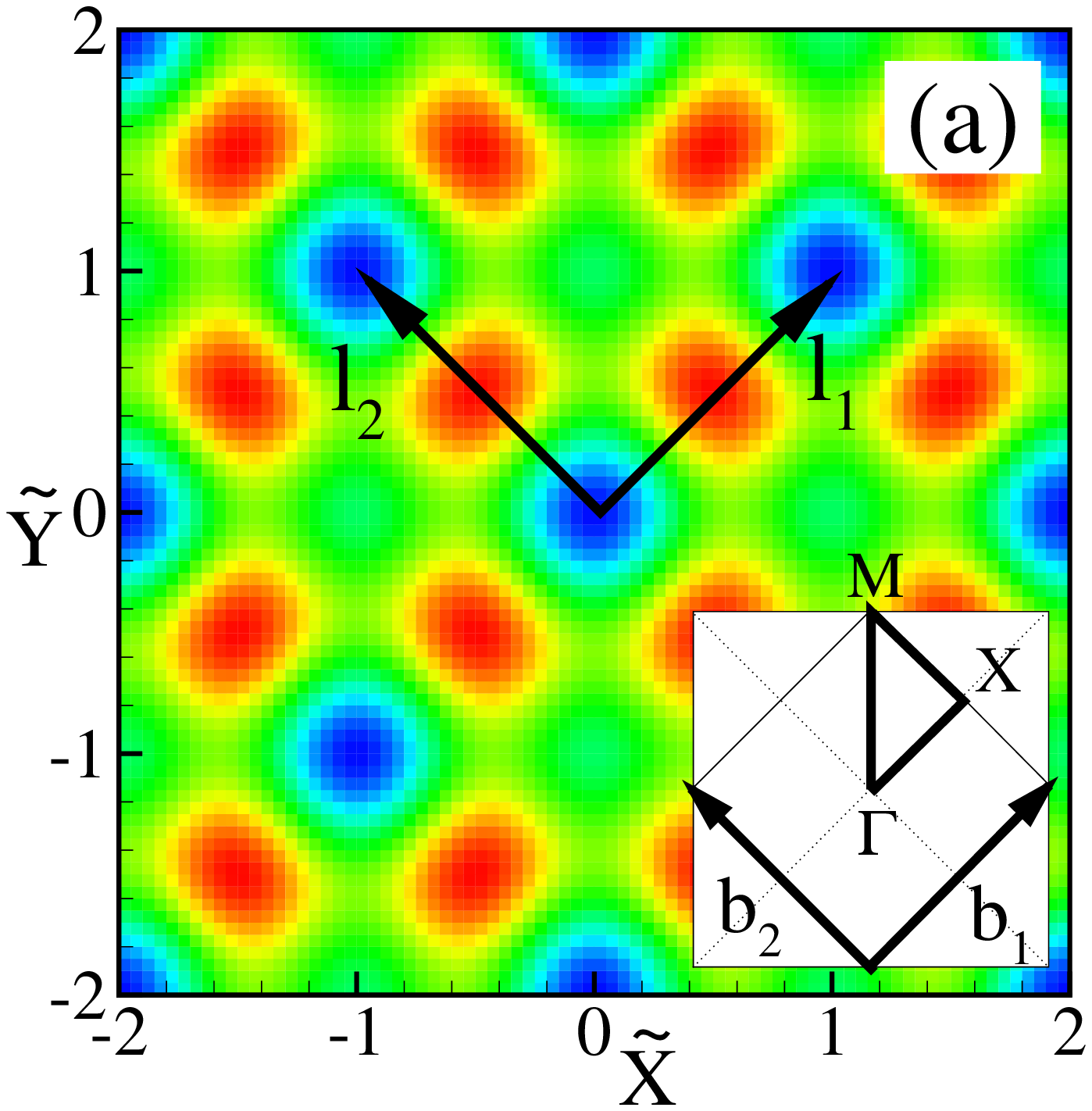,width=3.7cm}
\epsfig{file=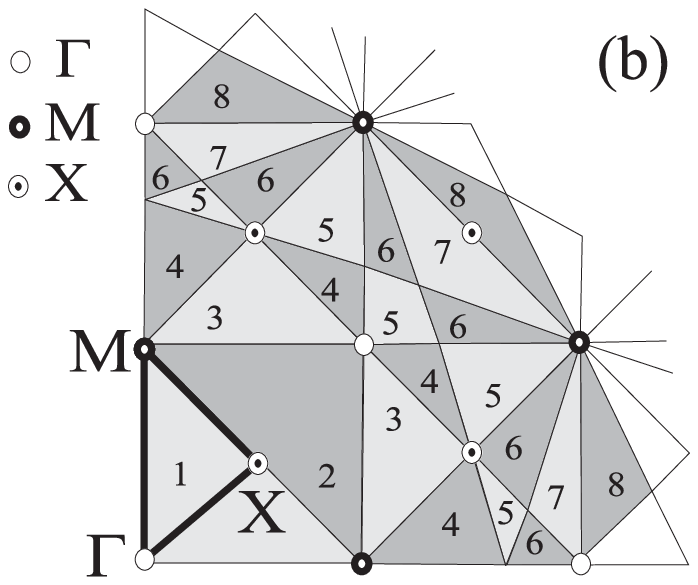,width=4.3cm}
\epsfig{file=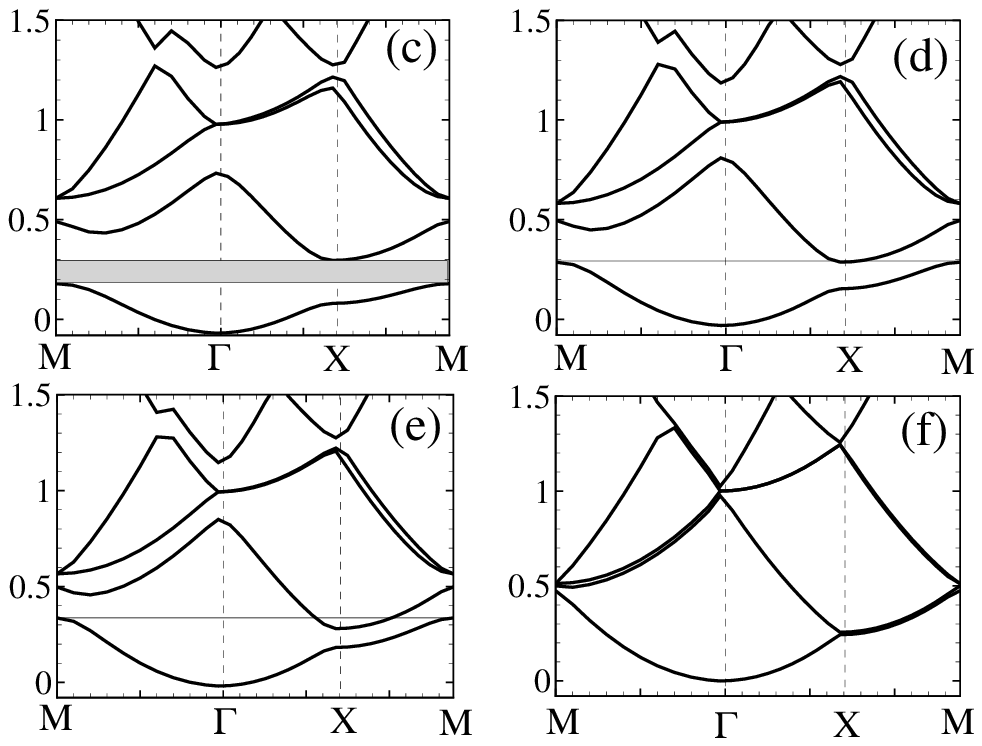,width=8cm}
\caption{ (Color online) (a) Contour plot of the potential
given by Eq. (\ref{potential}) with $V_0 = -0.55$ and $\varepsilon = 0.25$. In the inset
we display the primitive vectors ${\bf b}_{1,2}$  of the  reciprocal lattice, as
well as  the first BZ with the high symmetry points $\Gamma$, X, and M. (b) The eight lowest BZs with positions of high-symmetry points (only the first quadrant is shown).
(c-f) Structures of the $E(q)$ for different amplitudes
such as in (c) $V_0=-0.55$ (full gap),  in (d) $V_0=-0.3823$ (vanishing gap),
in (e) $V_0=-0.3$ (overlapping of the first and second bands), in (f)
$V_0=-0.05$ (shallow lattice).}
\label{fig0}
\end{figure}

It turns out, however that the laboratory system of coordinates, where $\tbR$ is
the radius-vector,  is not the most convenient frame for development of the
theory. This stems from the fact that a nonseparable lattice
possesses an elementary cell with boundaries that are rotated with respect to the axes of
the laboratory system. This is
illustrated in Fig.~\ref{fig0} where we show the OL given by
(\ref{potential}). The simple square lattice has the period $\sqrt 2 \pi$ which is
rotated by the angle $\pi/4$ with respect to the laboratory system. In Fig.~\ref{fig0}~(a)
besides the primitive translation vectors ${\bf l}_1 = \sqrt{2}\pi(1,1)$ and
${\bf l}_2 =\sqrt{2}\pi(-1,1)$, we depict in the inset the basis vectors
${\bf b}_1 = 1/\sqrt{2}(1,1)$ and ${\bf b_2} = 1/\sqrt{2}(-1,1)$ of the reciprocal
lattice which define the first Brillouin zone (BZ).

Therefore it is convenient to introduce the rotated coordinate
system~\cite{Kolovsky} defining $X=\tilde{X}+\tilde{Y}$ and $Y=\tilde{X}-\tilde{Y}$. In the new coordinates the potential (\ref{potential}) reads as
\begin{eqnarray}
\label{potential_rot}
V({\bf R}) = 2\varepsilon V_0  \left(\cos X + \cos Y +  \varepsilon^{-1} \cos X \cos Y\right)
\end{eqnarray}
and represents a simple square lattice with the primitive translation vectors
parallel to the rotated X and Y axes. In the new frame the GP equation reads as
\begin{eqnarray}
\label{GPE_gen}
i \partial_T\Psi =-2\Delta_{\bf R} \Psi+({\bf a}'\cdot{\bf R})\Psi+V(\bR) \Psi+\sigma  |\Psi|^2\Psi.
\end{eqnarray}
where ${\bf a}^\prime=\left((a_x+a_y)/2,(a_x-a_y)/2\right)$.

In the linear case inter-band tunneling represents a possibility for a particle to pass
from one energetic band to another. In 1D lattice this may  occur due to the linear
potential (the LZ tunneling), which can be interpreted in terms of changing the slope of
the gap edges, leading to equal "effective" energies of the initial and the final steps
(see e.g.~\cite{Ziman}). In 2D and 3D lattices a particle can
tunnel even without applying a linear force: in particular, in the case of a closed gap
when initial and final states satisfy the energy and quasi-momentum conservation laws:
\begin{eqnarray}
\label{reson_cond}
E_1=E_2\quad\mbox{and}\quad \bq_1-\bq_2=\bQ,
\end{eqnarray}
hereafter $\bQ$ is an arbitrary vector of the reciprocal lattice.

Even weak nonlinearity changes the situation dramatically. First, it can couple in a resonant way two (or more) states,
which do not satisfy the energy and momentum conservation laws
(\ref{reson_cond}). This occurs due
to process of {\it four-wave mixing}, which is efficient between the modes  possessing
the same group velocity. Second, the nonlinearity results in distinct stability
properties of the different states, which in particular lead to asymmetric
tunneling, i.e. to dependence of the tunneling on the initial
conditions~\cite{KKS,BKK}. We therefore use the term the {\em nonlinear tunneling} in
order to highlight essentially new features of the phenomenon.

In this context we recall that accounting spatial dependence of the models describing tunneling becomes highly relevant even in one-dimensional case where instabilities and nonexponential decay rate take place (see Refs. \cite{KKS} and \cite{Wimb,decay}, respectively).

\section{Finite-mode  approximation. Deep lattices.}
\label{sec:deep}

\subsection{Basic notations.}

The theory developed in the present section is restricted neither by the cos-like
potential as that of considered in (\ref{GPE}) nor by the 2D case. We, however, limit
ourselves only to the 2D case (3D generalization is straightforward but more cumbersome)
and, therefore,
by choosing the geometry of the square lattice, we first consider Eq. (\ref{GPE_gen}) with
a more general periodic potential $V(\bR)$: $V(\bR)=V(\bR+ {\bf l}_j)$, where ${\bf
l}_{j}$ ($j=1,2$) are the orthonormal lattice vectors.  For the analysis of
particular cases, we, however, always refer to the specific equation
(\ref{GPE}), (\ref{potential}).

We hereafter consider the lattice to be large enough in each direction,
what allows us to follow the standard practice to consider effects independent on
boundary conditions by choosing the most convenient ones, specifically we impose
the cyclic boundary conditions.

In the analytical part of the present paper we deal only with the rotated system $\bR$
considering $\Delta_\bR \equiv \nabla^2$ where $\nabla\equiv(\partial_X,\partial_Y)$.
We also use the derivative with respect to other variables indicating them  by subindices,
e.g. $\nabla_\bq=(\partial_{q_X},\partial_{q_Y})$.

In the weakly nonlinear case the associated linear spectral problem
\begin{eqnarray}
\label{cL}
    \cL\varphi_{\alpha \bq}(\bR)=E_\alpha (\bq)\varphi_{\alpha \bq}(\bR),\qquad {\cal L}\equiv -2\nabla^2+V(\bR)
\end{eqnarray}
plays a prominent role. Here
$\bq$ is the wavevector in the reduced BZ: $q_{X,Y}\in
[-q_B,q_B]$ with $q_B=1/2$, $\alpha=1,2,...$ is a band number,
$\varphi_{\alpha \bq}(\bR)=e^{i\bq\bR}u_{\alpha \bq}(\bR)$,
where $u_{\alpha \bq}(\bR)$ are periodic functions $u_{\alpha \bq}(\bR)=u_{\alpha
\bq}(\bR+{\bf l}_j)$, are the Bloch functions (BFs).

We use the internal product defined by
\begin{eqnarray}
\label{multiscale2}
 \langle u\mid \hat{A}\mid v\rangle =  \frac{1}{{\cal V}} \int_{{\cal V}}  \bar{u} (\br)\hat{A} v(\br)d\br,
\end{eqnarray}
where $\hat{A}$ is an operator and ${\cal V}$ is the volume of the lattice.
With the linear eigenvalue problem (\ref{cL}) we associate physically important quantities: the group
velocity of a mode
\begin{eqnarray}
\label{vel_disc}
    \bv_{\alpha\bq}= \langle \varphi_{\alpha\bq}\mid -4i \nabla\mid \varphi_{\alpha\bq}\rangle=\nabla_{\bq}E_\alpha(\bq)
\end{eqnarray}
and the tensor of the inverse effective mass $\bM_{\alpha\bq}^{-1}$  with the entrees
\begin{eqnarray}
\label{mass}
    (\bM_{\alpha\bq}^{-1})_{ij}= 4-8\sum_{\alpha_1\neq\alpha}\frac{ \langle
    \varphi_{\alpha\bq}\mid \partial_k \mid\varphi_{\alpha_1\bq}\rangle\langle \varphi_{
    \alpha_1\bq}\mid \partial_l \mid\varphi_{\alpha \bq}\rangle}{E-E_{\alpha_1}(\bq)}
    \nonumber \\
    =\frac{\partial^2E_\alpha(\bq)}{\partial q_i\partial q_j },
\end{eqnarray}
where $k,l=X, Y$ (see e.g. \cite{BKS} for more details).

In the case where the lattice potential is not small, i.e. $V_0\gtrsim 1$,
we seek the solution of Eq. (\ref{GPE_gen}) in the form
\begin{eqnarray}
\label{expan1}
    \Psi=\sum_{j=1}\epsilon^j\psi_j,
\end{eqnarray}
where $\epsilon\ll 1$ is a small parameter and $\psi_j$ are the functions of a
scaled spatial, ${\bf r}_j=(x_j,y_j)=(\epsilon^j X, \epsilon^j Y)$, and temporal, $t_j=\epsilon^j T$,
variables (here $j=0,1,...$).
By analogy we define  $\nabla_j=\left(\partial_{x_j},
\partial_{y_j}\right)$. This  approach is referred to as a multiple-scale expansion, which in the
context of 2D BEC loaded in an OL was described in~\cite{BKS}. Below we explore some
generalization of the expansion,  including the linear force as
well as resonant wave interactions in the consideration.

By substituting expansion (\ref{expan1}) into Eq. (\ref{GPE_gen}), passing to the scaled
independent variables, and gathering the terms of the same order of $\epsilon$, up
to $\epsilon^3$ one obtains ($j=1,2,3$)
\begin{eqnarray}
\label{eigenvalue}
\label{Cauchy}
    \left(i\partial_{t_0}-\cL\right)\psi_j={\cal F}_j,
\end{eqnarray}
where $
%&&
\cF_1=0$, $\cF_2=-(i\partial_{t_1}+4\nabla_0\nabla_1)\psi_1,$ and
%\\
\begin{eqnarray*}
&&
\cF_3=-(i\partial_{t_1}+4\nabla_0\nabla_1)\psi_2 -(i\partial_{t_2}+4\nabla_0\nabla_2+2\nabla_1^2)\psi_1
\\
&& \qquad
+\sigma|\psi_1|^2\psi_1.
\end{eqnarray*}

Tunneling of atoms between two states, belonging to the bands
$\alpha_1$ and $\alpha_2$  requires equality of the group
velocities of the interacting modes (see e.g.~\cite{Newell}):
\begin{eqnarray}
\label{velocity}
    \bv_1=\bv_2=\bv
\end{eqnarray}
what will be assumed in what follows.

Finally, to shorten notations, the BFs corresponding to the states coupled by tunneling
will be denoted as ($j=1,2,3$) $\varphi_{\alpha_j\bq_j}=\varphi_j$ (respectively $u_{\alpha_j\bq_j}=u_j$).

\subsection{Band structures}

Focusing on the lowest bands, in
Fig.~\ref{fig0} we show four qualitatively different configurations, which can be
realized by varying the amplitude of the
potential. They correspond to the existence of the full gap between the first
and second lowest bands, indicated by shaded area in the panel (c); closed (or
infinitesimally small) gap, indicated by the thin horizontal line in the panel (d);
absence of a full gap in a relatively deep lattice shown in the panel (e); and
the band structure associated with a shallow lattice depicted in the panel (f).

Different aspects of instabilities and inter-band transitions corresponding to  the
cases (c) of a full gap and (e) of absence of a gap, when resonance conditions of
the four wave mixing are not satisfied, have been considered in \cite{BKS} and
\cite{BKK}, respectively, and they will not be addressed here. We only note that
before instabilities are developed the problem is well described within the
framework of the two-level model.

\subsection{Nonlinear inter-- and intra--band tunneling.}
\label{intra-band}

Let us start with the case of the {\em resonant} nonlinear tunneling between two different
states having the same energy, i.e. satisfying the conditions
\begin{eqnarray}
\label{constrain1_reson}
    E_1=E_2=E\quad\mbox{and}\quad
     \bq_1 = \bq_2+\bq_2-\bq_1+ \bQ.
\end{eqnarray}
Both (\ref{constrain1_reson}) and (\ref{velocity}) can always be achieved in all orthogonal lattices if one considers
two X points related by the rotation by $\pi/2$: then $\bq_1=(q_B,0)$ and $\bq_2=(0,q_B)$;
and in lattices with a closed gap as it is shown in Fig.~\ref{fig0}~(d): then $\bq_1=(q_B,0)$
and $\bq_2=(q_B,q_B)$ and the points X and M are considered. The most significant difference between the mentioned
two cases is that in the first case the effective masses of the both states have the same sign
(and thus the respective Bloch states posses the same stability), while in the second case the
effective masses have different signs (and one of the Bloch states is necessarily stable, while
the other one is unstable).

We look for a solution of (\ref{eigenvalue}) with $j=1$ in the form
\begin{eqnarray}
\label{psi1_1}
    \psi_1=
    \left[
    A_1(\br_1,t_1)\varphi_1(\br_0)
    +
     A_2(\br_1,t_1)\varphi_2(\br_0)
     \right]
     e^{-iE t_0},
\end{eqnarray}
where $A_j(x_1,t_1)$ are slowly varying amplitudes which depend on slow variables
with the fastest ones being indicated explicitly. For the next step one expands (see
Appendix~\ref{ap:reson})
\begin{eqnarray}
\label{psi2_2}
    \psi_2
    &=&
    \left[
    \sum_{\alpha\neq\alpha_1}B_{\alpha}^{(1)}(\br_1,t_1)\varphi_{\alpha\bq_1} (\br_0)
    \right.
    \nonumber \\
    &+&
    \left.
    \sum_{\alpha\neq\alpha_2}B_{\alpha}^{(2)}(\br_1,t_1)
    \varphi_{\alpha\bq_2}(\br_0)\right]e^{-iEt_0},
\end{eqnarray}
where $B_{\alpha\bq}(\br_1,t_1)$ are slowly varying functions (we use a convention
that all sums with respect to $\alpha$ are computed between $\alpha=1$ and
$\alpha=\infty$, with possible exclusions which are explicitly indicated. Substituting (\ref{psi2_2}) in
(\ref{eigenvalue}) with $j=2$ and projecting on the states $\varphi_{1,2}$
we obtain
\begin{eqnarray}
\label{A_1}
    A_1=A_1(\br_1-\bv t_1, t_2)\,\,\,\mbox{and}\,\,\, A_2=A_2(\br_1-\bv t_1, t_2)\,.
\end{eqnarray}
Projection on the rest of the states $\varphi_{\alpha\bq}$ yields the expression for $B$
(see Appendix~\ref{ap:reson}).

Passing to the third order we impose the orthogonality of $\cF_3$ and
$\psi_1$ and
taking into account that the conditions (\ref{velocity}) and
(\ref{constrain1_reson}) can be satisfied simultaneously only for high-symmetric points,
e.g. the points $\Gamma$, X, and M, where the group velocity is zero, $\bv=0$ and
the tensor of the inverse effective mass is diagonal,
$\hat{\bM}_\alpha^{-1}=$diag$(M_{\alpha,x}^{-1},M_{\alpha,y}^{-1})$.  Then we obtain set of the
equations
\begin{subequations}
\label{deep_latt_reson_inter}
\begin{eqnarray}
i\frac{\partial A_1}{\partial t_2}&+&  \frac12 \nabla_1\hat{\bM}_1^{-1}\nabla_1 A_1
 -\tilde{\chi}_{12}\bar{A}_1A_2^2
 \nonumber \\
 &-&\left(\chi_{11}|A_1|^2+2\chi_{12}|A_2|^2\right)A_1=0,
    \\
i\frac{\partial A_2}{\partial t_2}&+&  \frac12 \nabla_1\hat{\bM}_2^{-1}\nabla_1
A_2-\tilde{\chi}_{21}\bar{A}_2A_1^2
    \nonumber\\
     &-&\left(\chi_{22}|A_2|^2+2\chi_{12}|A_1|^2\right)A_2 =0,
\end{eqnarray}
\end{subequations}
where we define the operator
\begin{eqnarray*}
\nabla_1\hat{\bM}_\alpha^{-1}\nabla_1 \equiv
    \frac{1}{M_{\alpha,x}}\frac{\partial^2}{\partial x_1^2}+\frac{1}{M_{\alpha,y}}
\frac{\partial^2}{\partial y_1^2}.
\end{eqnarray*}
The nonlinearity coefficients are given by ($\alpha,\beta=1,2$)
\begin{eqnarray}
\label{chi}
    &&\chi_{\alpha\beta}=\chi_{\beta\alpha}=\sigma \langle \varphi_\alpha \varphi_\beta\mid
\varphi_\alpha \varphi_\beta \rangle,
\\
\label{t_chi}
    &&\tilde{\chi}_{\alpha\beta}=\overline{\tilde{\chi}}_{\beta\alpha}=\sigma\langle \varphi_\alpha^2\mid
\varphi_\beta^2 \rangle.
\end{eqnarray}

\subsection{Nonlinear Landau-Zener tunneling.}

The LZ tunneling is induced by the linear potential when there exists a full gap
between bands. Being originated by the linear term it requires conservation of the linear momentum. In order to observe how
the nonlinearity affects LZ tunneling, the linear force has to induce an effect of
the same order as the nonlinearity. Since our approach is valid only for weak
nonlinearity, both the linear force and the gap, through which tunneling is
considered, have to be sufficiently small, as well. Thus we require
\begin{eqnarray}
\label{condLZ}
    E_2-E_1=\epsilon^2{\cal E} \,,\quad \bq_1-\bq_2 = \bQ,
\end{eqnarray}
where ${\cal E}\lesssim 1$.
Further, we assume
that ${\bf a}'=\epsilon^2 \widetilde{\ba}$ with $|\wa|\sim 1$. Therefore, the linear force will appear
only in the third order of the multiple-scale expansion.

The condition (\ref{condLZ}) together with the constraint (\ref{velocity}) means that in a general
situation LZ tunneling can occur either between the two symmetric points M, with $\bq_1=(q_B,q_B)$
and $\bq_2=(\pm q_B,\pm q_B)$, or between the points X, with $\bq_1=(q_B,0)$ and $\bq_2=(0,\pm q_B)$,
belonging to different boundaries of a gap. Therefore, as before the group velocity is zero, i.e. $\bv=0$.

Taking into account that now the energies of the two states are different, we perform the
multiple-scale expansion for the superposition of the states
\begin{equation}
\label{psi1_2}
    \psi_1=
    A_1(\br_1,t_1)\varphi_1(\br_0) e^{-iE_1 t_0}
    +
     A_2(\br_1,t_1)\varphi_2(\br_0)
     e^{-iE_2 t_0}
\end{equation}
and repeat the steps of derivation of (\ref{deep_latt_reson_inter})
(see Appendix~\ref{ap:multiscale}1). Specifically, the second order term now has the form (here $\bq=\bq_1$)
\begin{eqnarray}
\label{psi2_1_ap}
    \psi_2=\sum_{\alpha}\left[B_{\alpha\bq}^{(1)}(\br_1,t_1)\varphi_{\alpha\bq}e^{-iE_1t_0}
    \right.
    \nonumber
    \\
     +
     \left.
     B_{\alpha\bq}^{(2)}(\br_1,t_1)\varphi_{\alpha\bq}e^{-iE_2t_0}\right].
\end{eqnarray}
 The requirement of the absence of the secular terms results in
\begin{subequations}
\label{deep_latt_nonreson_inter}
\begin{eqnarray}
    i\frac{\partial A_1}{\partial t_2}&+& \frac12  \nabla_1\hat{\bM}_1^{-1}\nabla_1 A_1
-\kappa   e^{i{\cal E}t_2}A_2 \nonumber\\
&-&\left(\chi_{11}|A_1|^2+2\chi_{12}|A_2|^2\right)A_1 =0,
    \\
 i\frac{\partial A_2}{\partial t_2}&+& \frac12 \nabla_1\hat{\bM}_2^{-1}\nabla_1 A_2
    -\kappa   e^{-i{\cal E}t_2}A_1 \nonumber\\
  &-&\left(\chi_{22}|A_2|^2+2\chi_{12}|A_1|^2\right)A_2 =0,
\end{eqnarray}
\end{subequations}
where the coupling coefficient
$\kappa$ is given by
\begin{eqnarray}
\label{kappa}
    \kappa=\wa\langle\varphi_1|\br|\varphi_2\rangle=\wa\langle\varphi_2|\br|\varphi_1\rangle\,.
\end{eqnarray}
In the case, when the Rabi frequency defined by $\kappa$
significantly exceeds the rotational frequency ${\cal E}$,  i.e. when ${\cal E}\ll \kappa$, the time
dependence of the coupling term can be neglected.

\subsection{Interplay between Landau-Zener tunneling and nonlinear tunneling.}

In the cases considered above we singled out one resonant process, which is
dominant for the atomic migration between the bands. Meantime, one can create a
situation where there exist two competing processes. This, in particular, may happen
in the structure, depicted in Fig.\ref{fig0}~(d), when a linear force is applied allowing
for resonant nonlinear LZ tunneling. In this particular case, the state M of the lowest band
(state "1" below) through the linear  force is coupled to the state M of the second band
(state "2" below) and through the four-wave mixing process with the state X of the second
band (state "3" below). The matching conditions now read
\begin{subequations}
\begin{eqnarray}
\label{constrain1_nonresonant_intra}
    && E_1=E_3=E_2-\epsilon^2{\cal E},
    \\
     && \bq_1 =\bq_2+ \bQ, \quad \bq_1 = \bq_3-\bq_3+\bq_1+ \bQ.
\end{eqnarray}
\end{subequations}

 To take into account all resonant processes in this case a three-mode model is required.
Therefore, we seek a solution of Eq. (\ref{eigenvalue}) in the form
\begin{eqnarray}
\label{psi1_3}
    \psi_1=
    [A_1(\br_1,t_1)\varphi_1(\br_0)+A_3(\br_1,t_1)\varphi_3(\br_0)] e^{-iE_1 t_0}
    \nonumber \\
    +
     A_2(\br_1,t_1)\varphi_2(\br_0)
     e^{-iE_2 t_0}.
\end{eqnarray}

The algebra similar to one described above, results in the following set of equations
for slowly varying amplitudes
\begin{widetext}
\begin{subequations}
\label{deep_latt_nonreson_inter_2}
\begin{eqnarray}
  &&  i\frac{\partial A_1}{\partial t_2}+ \frac12  \nabla_1\hat{\bM}_1^{-1}\nabla_1 A_1
-\kappa  e^{i{\cal E}t_2} A_2
\nonumber\\ &&
-\left(\chi_{11}|A_1|^2+2\chi_{12}|A_2|^2+2\chi_{13}|A_3|^2\right)A_1
-\tilde{\chi}_{13} \bar{A}_1A_3^2 =0,
    \\
&&   i\frac{\partial A_2}{\partial t_2}+ \frac12 \nabla_1\hat{\bM}_2^{-1}\nabla_1 A_2
    -\kappa  e^{-i{\cal E}t_2} A_1
    %\nonumber\\    &&
  -\left(\chi_{22}|A_2|^2+2\chi_{12}|A_1|^2+2\chi_{13}|A_3|^2\right)A_2 =0,
     \\
 &&   i\frac{\partial A_3}{\partial t_2}+ \frac12 \nabla_1\hat{\bM}_3^{-1}\nabla_1 A_3
    %\nonumber\\   &&
  -\left(\chi_{22}|A_2|^2+2\chi_{12}|A_1|^2+2\chi_{13}|A_3|^2\right)A_3-\tilde{\chi}_{31} \bar{A}_3A_1^2 =0.
\end{eqnarray}
\end{subequations}
\end{widetext}

We mention that resonant coupling of  three states (the two X states and one M
state) is also possible in the case depicted Fig.\ref{fig0}~(d). Detailed study of
the coupled-mode equations for the slow envelopes can be found in \cite{21}.

\section{Finite-mode approximation. Shallow lattice.}
\label{sec:shallow}

The LZ tunneling in a shallow accelerated lattice, $V_0\ll 1$, can be most
conveniently considered in the rotated coordinate system, i.e.  Eq. (\ref{GPE_gen})
with the potential defined by Eq. (\ref{potential_rot}).  For the sake of
convenience, we normalize the wave function as $\Psi=\psi/N^{1/2}$, and introduce
the nonlinearity coefficient $g = \sigma N$, with the assumption that $|g|\ll 1$.
Now the BFs can be approximated by   linear combinations of the plane waves (see
e.g. \cite{SHCH}), thus one can use the plane waves as the basis of expansion.
%This approach leads to the
%Landau-Zener-Majorana type models, which are derived below in this Section.

We assume that the order parameter of a BEC has  the Fourier spectrum consisting of
narrow peaks, whose width being much smaller than the size of the BZ. In this case,
similar to Houston's approach in the theory of accelerating electrons
\cite{Houston}, the order parameter is taken in the form
\begin{eqnarray}
\Psi(\bR,t) = e^{i\bq(t)\bR}\sum_{j}C_j(t)e^{i\bQ_j\bR},
\end{eqnarray}
where the sum consists of the plane waves satisfying the resonant conditions
(\ref{reson_cond}), $\bQ_j$ are the reciprocal lattice vectors.

Linear LZ tunneling is a result of Bragg resonance and occurs when the Bloch index
$\bq(t)$ crosses one of the Bragg planes (a Bragg plane in 2D is a line passing
through the mid-point of a reciprocal lattice vector $\bQ_j$, taken  from the
$\Gamma$-point, and perpendicular to it). The lattice defined by the potential
(\ref{potential_rot}) admits two-fold (i.e. quasi 1D) and four-fold Bragg
resonances (for more details see Ref. \cite{LZ2D}).

A weak nonlinearity can be accounted in the following way \cite{SHCH}: the lattice
potential and the nonlinear term are considered as an effective time-dependent
lattice potential $V_\mathrm{eff} \equiv V(\bR) + g|\Psi(\bR,t)|^2$, where the
order parameter is  assumed periodic in $\bR$. Thus, in this approach, nonlinearity
does not change the relation between LZ tunneling and Bragg resonances.

The potential can be rewritten in terms of the reciprocal lattice vectors  as
\eqb
V = \vare V_0[ e^{i\bQ_{1,0}\bR} + e^{i\bQ_{0,1}\bR} ] +
\frac{V_0}{2}[e^{i\bQ_{1,1}\bR} + e^{i\bQ_{1,-1}\bR}]+ c.c..
\label{POTQ}\eqe
Here $\bQ_{1,0} = (2q_B,0)$, $\bQ_{0,1} = (0,2q_B)$, and $\bQ_{1,\pm 1} =
\bQ_{1,0}\pm \bQ_{0,1}$ (recall that $q_B=1/2$). The Bragg planes can thus be
indexed accordingly, for instance, the Bragg plane $B_{1,0}$ corresponds to
$\bQ_{1,0}$.
\medskip

\noindent\textit{Two-fold Bragg resonances.} The two-fold Bragg resonance
occurs at any one of the Bragg planes away from the $\mathcal{O}(V)$-neighborhood
of the high-symmetry points, (i.e. the points of intersection of two or more Bragg
planes), for instance the M-points.  By keeping only the resonant terms we obtain:
\eqb
\Psi = C_1(t)e^{i\bq(t)\bR} + C_2(t)e^{i(\bq(t)-\bQ)\bR},
\label{EQ1}\eqe
\begin{eqnarray}
V_{\mathrm{eff}} = & & (\hat{V}_\bQ+ gC_1\bar{C}_2)e^{i\bQ\bR} + (\hat{V}_{-\bQ}+
g\bar{C}_1C_2)e^{-i\bQ\bR}\nonumber\\
& & + g(|C_1|^2 +|C_2|^2),
\label{EQ2}\end{eqnarray}
where $\hat{V}_\bQ$ is the Fourier amplitude corresponding to $\bQ$. Substituting
the expressions (\ref{EQ1}) and (\ref{EQ2}) into equation (\ref{GPE_gen}) we get
$\dot{\bq}=-\ba^\prime$, which cancels the linear potential, and the set of the
equations for the amplitudes $C_1$ and $C_2$. The latter can be simplified when we
use the conservation of the norm, in this case $|C_1|^2 +|C_2|^2=1$, and the phase
transformation $C_j = e^{-i\phi}c_j$ with $\dot{\phi} = 3g/2 +
2(\bq_{\mathrm{res}}^2 + (\ba^\prime)^2t^2)$, where we denote $\bq_\mathrm{res}$
the crossing point of the corresponding Bragg plane. Setting $\bq(t) =
\bq_\mathrm{res}-\ba^\prime t$ (i.e. the resonance is set at $t=0$), we obtain the
set of the equations ($j=1,2$)
\begin{equation}
i\dot{c_j}  =  (-1)^j\left[4\bq_\mathrm{res}\ba^\prime t + \frac{g}{2}(|c_j|^2 -
|c_{3-j}|^2)\right]c_j + \hat{V}_\bQ c_{3-j},\label{EQ3}
%\\
%i\dot{c_2} & = & \left[4\bq_\mathrm{res}\ba^\prime t + \frac{g}{2}(|c_1|^2 -
%|c_2|^2)\right]c_2 + \hat{V}_\bQ c_1.
%\label{EQ4}^
\end{equation}
The system   (\ref{EQ3})  is identical to that in the 1D lattice
\cite{Landau-Zener}. The only signature of the higher dimensionality in the system
(\ref{EQ3}) is that of the dependence of the sweep amplitude (i.e.
$4\bq_\mathrm{res}\ba^\prime$) on the angle between the Bloch index of the crossing
point and the acceleration.

In the linear case, $g=0$, the tunneling probability $P$ corresponding to the
system (\ref{EQ3}) is given by the well-known Landau-Zener-Majorana formula
\cite{LZM}:
\eqb
P \equiv |{c_1(\infty)}|^2 = \exp\left\{-\frac{\pi
|\hat{V}_\bQ|^2}{4|\bq_\mathrm{res}\ba^\prime|}\right\},
\label{EQ5}\eqe
where it is assumed that ``initially" $|c_1(-\infty)|=1$.  For instance, at the
X-point we get $P_{\rm X} = e^{-\pi\vare^2V_0^2/(a_x+a_y)}$.
\medskip

\noindent\textit{Four-fold Bragg resonances.} This resonance occurs at the
intersection of two Bragg planes, for instance at the M-point [the other case is at
the $\Gamma$-point, see Fig.\ref{fig0}~(f)]. Using the lattice potential in
 (\ref{POTQ}),   the resonant part of the order parameter
\eqb
\Psi_\mathrm{res} = \sum_{j=1}^4C_j(t)e^{i(\bq_j-\ba^\prime t)\bR},
\label{EQ7}\eqe
where $\bq_1 = \bQ_{1,1}/2$, $\bq_2 = \bQ_{-1,-1}/2$, $\bq_3 = \bQ_{-1,1}/2$, and
$\bq_4 = \bQ_{1,-1}/2$ connect the $\Gamma$ and M points, and the relations
$\bQ_{1,1}\ba^\prime = a_x$ and $\bQ_{1,-1}\ba^\prime = a_y$ we obtain a set of the
equations for the Fourier amplitudes $C_j$ of the order parameter. The system is
simplified after the transformation $C_j = e^{-i\phi}c_j$ with $\dot{\phi} = g +
2[\bq_1^2 + (\ba^\prime)^2t^2]$ and using of the norm conservation
$\sum_j|C_j|^2=1$:
\begin{widetext}
\begin{eqnarray}
i\dot{c}_1 &=& \left(-2a_x t - g|c_1|^2\right)c_1 + \frac{V_0}{2}c_2 + \vare
V_0(c_3+c_4) + 2g\bar{c}_2c_3c_4,\label{EQ9}\\
i\dot{c}_2 &=& \left(2a_x t - g|c_2|^2\right)c_2 + \frac{V_0}{2}c_1 + \vare
V_0(c_3+c_4) + 2g\bar{c}_1c_3c_4,\label{EQ10}\\
i\dot{c}_3 &=& \left(2a_y t - g|c_3|^2\right)c_3 + \frac{V_0}{2}c_4 + \vare
V_0(c_1+c_2) + 2gc_1c_2\bar{c}_4,\label{EQ11}\\
i\dot{c}_4 &=& \left(-2a_y t - g|c_4|^2\right)c_4 + \frac{V_0}{2}c_3 + \vare V_0
(c_1+c_2) + 2gc_1c_2\bar{c}_3.\label{EQ12}
\end{eqnarray}
\end{widetext}

In the linear case, $g=0$, one can use the general formula for the tunneling
probability \cite{BE}, $P = |c_j(\infty)|^2= \exp\left\{-2\pi\sum_{m\ne j}
\frac{|\Delta_{mj}|^2}{|\nu_m-\nu_j|}\right\}$ for $|c_j(-\infty)|=1$, where
$\nu_j$ is the strongest sweep amplitude and $\Delta_{mj}$ is the cross-coupling
term. For instance, in the case of $|a_x|>|a_y|$ we get for $P = |c_1(\infty)|^2$:
\eqb
P =\exp\left\{-\frac{\pi V_0^2}{8|a_x|} -\frac{\pi
\vare^2V_0^2}{|a_x+a_y|}-\frac{\pi \vare^2V_0^2}{|a_x-a_y|}\right\},
\label{EQ13}\eqe
with the initial condition $|c_1(-\infty)|=1$. The same formula is true for
$P=|c_4(\infty)|^2$ with $|c_4(-\infty)|=1$. The physical meaning of Eq.
(\ref{EQ13}) is clear: it determines the fraction of  BEC  atoms which pass through
the border of the first BZ at the M-point by continuously increasing their
quasimomentum (BEC atoms leave the first BZ also through the side M-points, with
the corresponding fractions given by the Fourier amplitudes $c_3$ and $c_4$).

In the special case of $a_x = a_y = a$ using formula (\ref{EQ5}), we obtain the
probability of the transition  in the form
\eqb
P = \exp\left\{-\pi\frac{(1/2+\vare)^2V^2_0}{2|a|}\right\}.
\label{EQ16}\eqe

Finally, the  squared amplitudes $|c_j(t)|^2$ of the LZ system can be compared with
the powers of the Fourier peaks in the full PDE (defined as the integral of the
modulus squared of Fourier amplitude) -- see Fig. \ref{Zener1} of Section
\ref{sec:numshallow}.

\section{Numerical results}
\label{numerics}

The theory of tunneling developed in the preceding sections is based on the two-- (few --)
mode approximation. As it has been shown in~\cite{BKK} on an example of  inter-band tunneling
between two initially populated states, this approach is valid until instability is developed.
In general, when initially only one of the states is populated or when one assumes several
populated states, that are not connected by some of the resonant conditions, the theoretical
model requires verification.

In this Section we address  problems of tunneling starting
with  only one populated state in a stationary lattice and of LZ tunneling in an accelerating
lattice with different directions of the lattice acceleration and different lattice depths. We impose the initial conditions in the form
$
%\begin{equation}
\psi(x,0)=\sum_{\alpha,\bq }\sqrt{r_{\alpha\bq}}\, \varphi_{\alpha\bq},
%\end{equation}
$
where summation is taken up to the eight lowest bands, $ \alpha=1,...,8 $, and over the symmetric
points which we denote as
$\bq_\Gamma=(0,0)$, $\bq_{\rm X}=(\pm 1/2,\pm 1/2)$, $\bq_{\rm X^\prime}=(\pm 1/2,\mp 1/2)$, and
$\bq_{\rm M}=[(0,\pm 1),(\pm 1,0)]$. We study evolution of the positive quantities $r_{\alpha\bq}$
representing populations of the respective states. Also we use the notation Y$_\alpha$ for a point Y=(X, M, $\Gamma$)
of $\alpha$-th band.

In all numerical simulations we solve the original GP  equation (\ref{GPEa}), i.e.
our results are obtained directly for the accelerating lattice  (\ref{OL_accel}).
We use the Crank-Nicolson  scheme with periodic boundary conditions  on the grid
with $400\times400$ points and with 20 periods of the OL in each direction (such a
size of the lattice lattice is sufficiently large to allow one to observe volume
phenomena that are not affected by the specific type of the boundary conditions).
Finally, we assume inter-atomic interactions to be repulsive, i.e. $\sigma=1$.

For a typical experimental setup with $10^4$ atoms having  scattering length of
order of $0.01$~nm (achievable by means of Feshbach resonance) and with transversal
size of the cloud $\ell_z\approx 0.3\,\mu$m our numerical unit of time corresponds
approximately to $0.25$~ms. Also, in what follows we use the normalization
constants $N=10$ and $N=20$ [defined after (\ref{GPEa})] what corresponds to
$3\times 10^4$ and $6\times 10^4$ of real atoms.

\subsection{Stationary optical lattice}
\label{subsec:stationary}

%\subsubsection{Modulational instability}
\subsubsection{Non-tunneling regime and the modulational instability}

\begin{figure}[h]
\epsfig{file=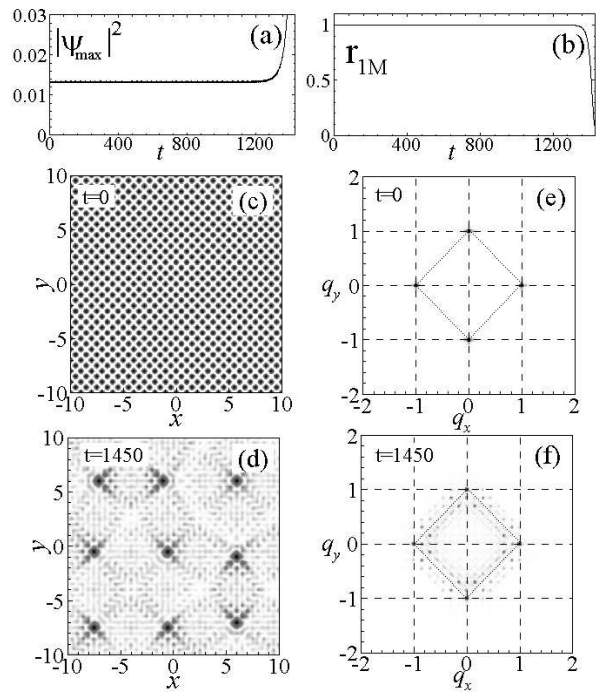,width=8cm,angle=0}
\caption{(Color online) Development of MI starting in the M$_1$-point.
In (a) temporal behavior of the density maximum [at $t\approx 1300$ (what corresponds 0.325~s, i.e. to experimentally feasible time)
the MI starts to develop] and in (b) the
dynamics of the population of the M$_1$-state is shown (populations of the higher bands
in the $\Gamma$\mbox{-,} X- and M-points are smaller than $10^{-5}$).
The parameters are as follows: $V_0=-0.3823$, $\varepsilon=0.25$,
$N=10$.
In (c), (d) density profiles  and in (e), (f) their Fourier transformations
at different times  are shown.}
\label{fig_nln_M1}
\end{figure}

We start with examining the case where energy of the initially populated
M$_1$-point is the same as that of X$_2$-point which is not populated at $t=0$.
This is the case described by  (\ref{constrain1_reson}). Now in contrast to the
situation described in~\cite{BKK}, now it is {\it a priori} not known which mode
(or modes) the energy will be transferred to. Thus one may expect tunneling between
bands, induced by modulational instability (MI) in the state M$_1$(since  the
effective masses $M_{1,x}$ and $M_{1,y}$ are negative  and  $\sigma=1$). 
The results are shown in Fig.~\ref{fig_nln_M1}.

The most relevant result stems from the fact that even after the MI is developed the populations of the symmetric states of the upper
bands, including the states X$_2$ are negligibly small. It means, that in spite of the resonant
conditions (\ref{constrain1_reson}) being satisfied, tunneling does not occur, i.e. the expected scenario,
which is described above, is not realized. In order to investigate the dynamics in the
{Fourier} space, we have computed the Fourier transform of the wave function at
different moments of time see Fig.~\ref{fig_nln_M1}~(e), (f).
 In the process of the development of MI the   atoms spread out along the boundary of the
first BZ. Dispersion spreading proceeds until the MI is developed. Only the unstable state M$_1$ imprints its signature in the symmetry of the developed
structure  [see Fig.~\ref{fig_nln_M1}~(d)] and therefore its pattern closely resembles the structures
emerged from the unstable states in a separable lattice (c.f. with Fig.~7 in the Ref.\cite{BKS}).

%\subsubsection{Intra-band tunneling}
\subsubsection{Bloch intra-band tunneling at the X$_2$-point}

\begin{figure}[h]
\epsfig{file=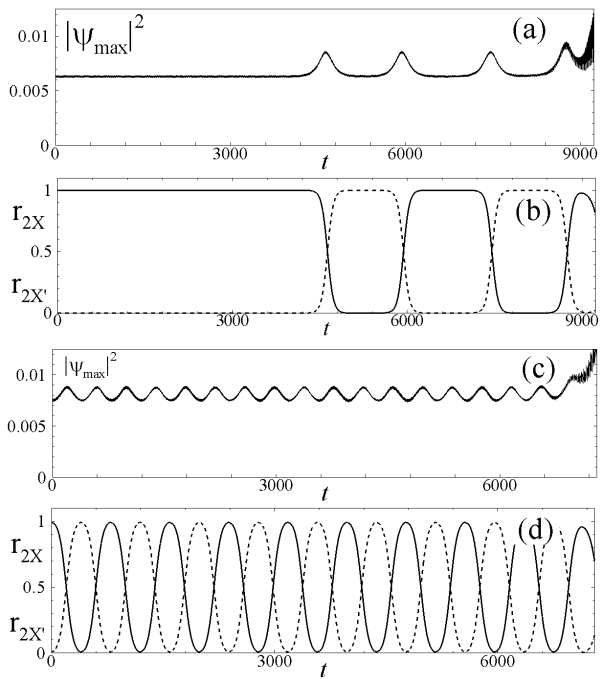,width=8cm,angle=0}
\caption{Time dependence of the maximum of the density amplitude [(a),(c)] and  populations
r$_{\rm 2X}$ (solid line) and r$_{\rm 2X^\prime}$ (dashed line) [(b),(d)] of the X-states of the second band
in the case when in (a), (b) initially all particles are in the X$_2$-state and in (c), (d) initially
r$_{\rm 2X}=0.99$ and r$_{\rm 2X^\prime}=0.01$.  In (a) at $t\approx 9000$ ($\approx 2.3$~s) and in (c) at $t\approx 7300$ ($\approx 1.8$~s)
the MI starts to develop. The ratio between the time intervals when the particles populate X- or
X$^\prime$-states in (b) is $\approx 1.2$. The parameters  are as follows: $\Lambda\approx 0.792$,
$V_0=-0.3823$, $\varepsilon=0.25$, $N=10$.% ($\approx 30000$ atoms).
}
\label{fig_nln_X2}
\end{figure}

\begin{figure}[h]
\epsfig{file=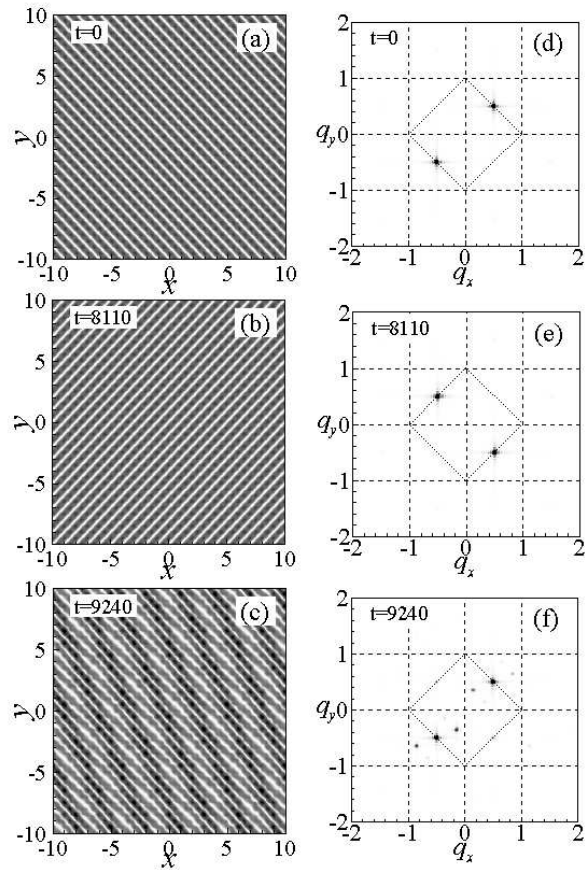,width=8cm,angle=0}
\caption{(Color online) In (a)-(c) density profiles  and in (d)-(f) their Fourier transformations
at different times corresponding to Fig.\ref{fig_nln_X2}~(a),~(b) are shown.}
\label{fig_nln_X21}
\end{figure}

We obtained perhaps even more surprising results
when in the lattice with the band structure depicted in Fig.~\ref{fig0}~(d) we initially populated a
stable state  X$_2$ that possesses two positive effective masses.  In Figs.~\ref{fig_nln_X2} and~\ref{fig_nln_X21}
one can  observe exchange of particles between the two doubly degenerate X$_2$-points which are related
by the rotation $\pi/2$ and referred to as X$_2$ and X$_2^\prime$ below. In other words the system
develops intra-band tunneling instead of the inter-band one.

This picture agrees with the theory developed in Sec.~\ref{intra-band}. Indeed, first we recall that
the both states X$_2$ are stable, while the state M$_1$ having the same energy is unstable. Next, we
consider (\ref{deep_latt_reson_inter}) and take into account that initially the envelopes
$A_{1,2}\equiv A_{1,2}(t_2)$ are independent on spatial coordinates: $\nabla_1A_{1,2}\equiv 0$.
This allows us to reduce the system of PDEs to a Hamiltonian system
with the Hamiltonian
\begin{eqnarray}
\label{ham_dyn_sys1}
    H_{eff}=\frac 12 \chi_{11}|A_1|^4+\frac 12 \chi_{11}|A_2|^4 +2\chi_{12}|A_1|^2|A_2|^2
    \nonumber \\
    +\chi_{12}\bar{A}_1^2A_2^2+\chi_{12}\bar{A}_2^2A_1^2\,.
\end{eqnarray}
Here we have taken into account the symmetry of the lattice resulting in
$\chi_{11}=\chi_{22}$ and $\chi_{12}=\tilde{\chi}_{12}=\chi_{21}=\tilde{\chi}_{21}$ [see Eqs. (\ref{chi}) and (\ref{t_chi})].

Since the derived dynamics preserves the "total number of particles" ${\cal N}=
{\cal N}_1+{\cal N}_2$, where ${\cal N}_j=|A_j|^2$ we introduce the population imbalance $z=(|A_2|^2-|A_1|^2)/{\cal N}$, the relative argument
$\phi=2\arg(A_1\bar{A_2})$, the parameter $\Lambda=\chi_{21}/\chi_{11}$, as well as the renormalized
time $\tau=\chi_{11}{\cal N}t_2$. Now the dynamical system  can be rewritten in the form
\begin{eqnarray}
\label{z}
    &&\frac{dz}{d\tau}=\Lambda (1-z^2)\sin\phi=\frac{\partial H_{eff}}{\partial \phi},
    \\
\label{phi}
    &&\frac{d\phi}{d\tau}=2z(1-\Lambda\cos\phi-2\Lambda)=-\frac{\partial H_{eff}}{\partial z},
\end{eqnarray}
where the effective Hamiltonian has the form
\begin{eqnarray}
\label{h_eff_1}
    H_{eff}=(1-z^2)(1-\Lambda\cos\phi-2\Lambda)\,.
\end{eqnarray}

For the next consideration one should keep in mind that the dependent variables are bounded:
$-1\leq z\leq 1$ and $0\leq \phi\leq 2\pi$. From Cauchy inequality
$\langle |\varphi_\alpha|^2|\varphi_\beta|^2 \rangle \leq \sqrt{\langle |\varphi_\alpha|^4\rangle\langle
|\varphi_\beta|^4\rangle}$, it follows that $0\leq \Lambda\leq 1$, what in the
case at hand is transformed into  $\chi_{12}^2\leq\chi_{11}^2$.

It is important to note here that the only  nonlinearity contribution  controls the
scaling of the time $\tau$, while the parameter $\Lambda$ is completely determined
by the lattice itself. Thus the nonlinearity sets the time scale, but not the type
of the dynamics (see below), which is defined by the lattice.

\begin{figure}[h]
\epsfig{file=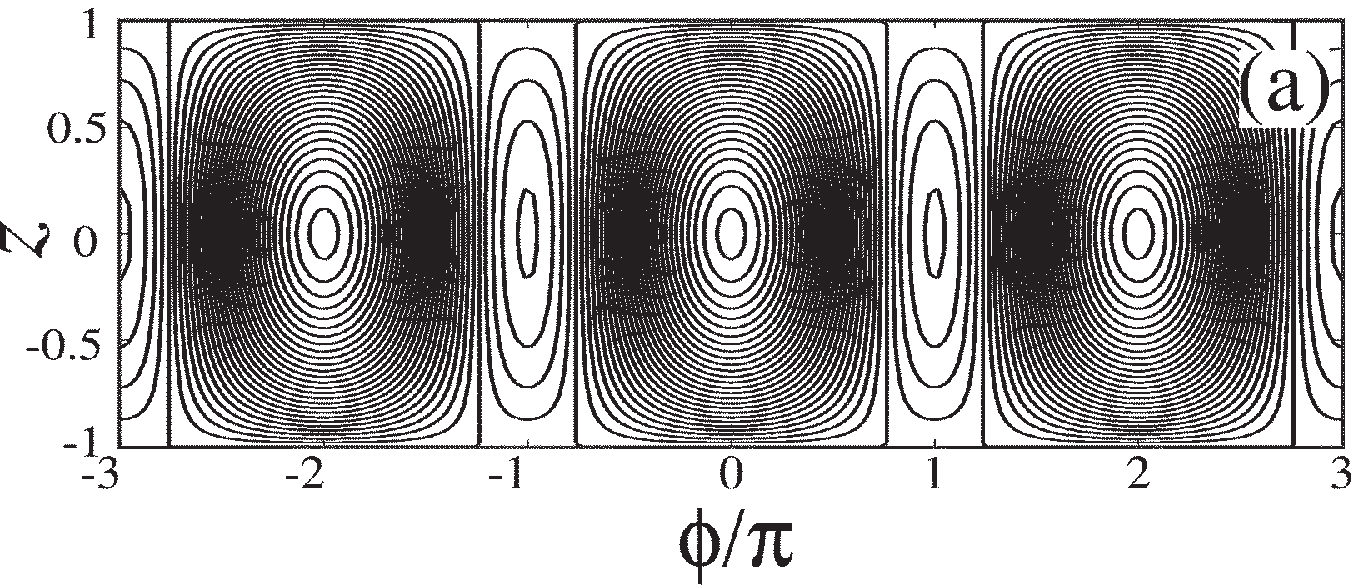,width=7cm,angle=0}
\epsfig{file=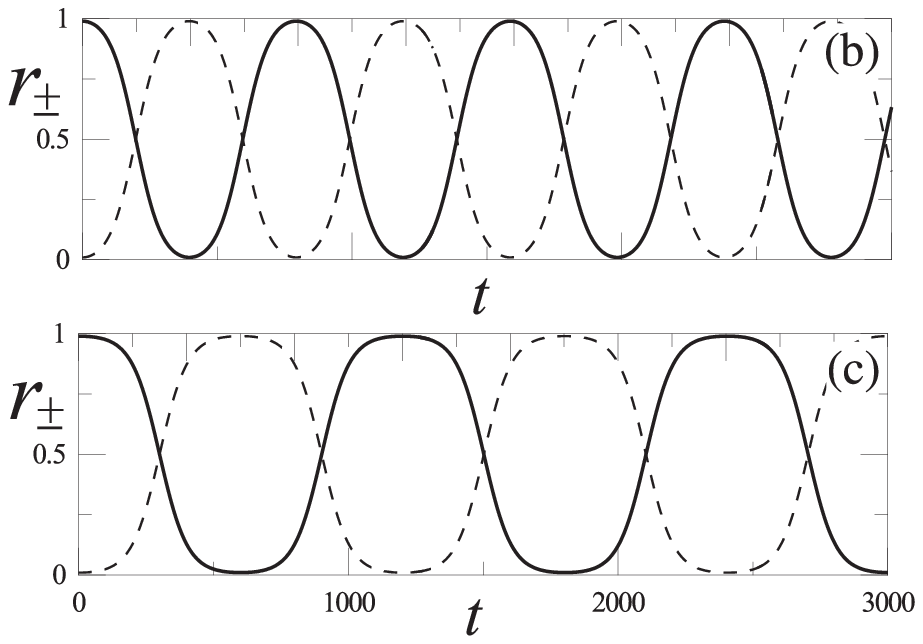,width=7cm,angle=0}
\caption{In (a) phase portrait of the system (\ref{z}), (\ref{phi}) with
$\Lambda=0.79 > \Lambda_c$ is shown.
In (b) and (c) the oscillations of the normalized population imbalance
with the initial conditions $z(0)= 0.98$, $\phi(0)=0$ in (b) and $\phi(0)=\pi$ in (c) correspondingly.
Here solid and dashed lines correspond to $r_+$ and $r_-$.
}
\label{fig4}
\end{figure}

The Hamiltonian (\ref{h_eff_1}) has two fixed points $F_1 = \{z=0,\phi = 0\}$ and $F_2 =
\{ z=0,\phi=\pi\}$: $F_1$ is a saddle point for
$\Lambda<1/3$ and a local minimum for $\Lambda>1/3$, while $F_2$ is always a local maximum
(since $\Lambda<1$). In Fig.~\ref{fig4}a we show the phase portrait corresponding to regime for
$\Lambda>\Lambda_{c}=1/3$.

In the panels (b) and (c) of Fig.\ref{fig4}  we  present the dynamics of the
"trajectories", $r_\pm (t)=[1\pm  z(t)]/2$, that correspond to the populations
$r_{2X}$ and $r_{2X^\prime}$ in the full picture illustrated by
Fig.~\ref{fig_nln_X2}. When all particles are initially in the X$_2$-state one
obtains $\{z(0)=1, \phi(0)=0\}$, what corresponds to a saddle point. This fact
explains why the initial dynamics of the population Fig.\ref{fig_nln_X2}~(b) is
constant. By time, however numerical errors accumulate what results in motion of a
system along the separatrix which separates two {adjacent} regions with different
dynamics. In the course of evolution the system having reached the next saddle
point jumps from one region to the other where the period of oscillations is
different. This process continues and it demonstrates itself through the
transitions in dynamics of the populations in Fig.\ref{fig_nln_X2}~(b). In order to
verify that the transitions occur due to the fact that the system stays close to
the separatrix we impose a small shift by taking initially r$_{2X}=0.99$ and
r$_{2X^\prime}=0.01$, which corresponds to  $\{z(0)= 0.98, \phi(0)=0\}$. As one can
see in Fig.~\ref{fig_nln_X2}~(d) the system immediately starts to oscillate with
the fixed period.  This can be interpreted as the oscillations around the fixed
point $F_1$ [c.f. Fig.~\ref{fig4}~(b)]. By comparing the ratio between the periods
in different regions near the separatrix, in  Fig.\ref{fig4}~(b), with the ratio
between different times in  Fig.\ref{fig_nln_X2}~(b), we found that they are 1.4
and 1.2 respectively.

Returning to Fig.~\ref{fig_nln_X2}, one can observe that the particle transfer due to the intra-band tunneling
is accompanied by small oscillations of the density amplitude. After some time the MI is developed. 
From the first sight this dynamics looks surprising because the energy interchange occurs between two stable
states. However, by inspecting the initial state in the points of high symmetry of the first
five bands we found that due to inherent inaccuracy of the numerical orthogonalization procedure initial data
have nonzero  contributions ($\sim 10^{-4}\div 10^{-5}$) from the unstable points of higher bands (for example,
in the case of Fig.\ref{fig4}~(a), (b) we have also contribution from the unstable points
r$_{\rm 3X^\prime}\approx 1.3\times 10^{-5}$ and r$_{\rm 5X}\approx 1.2\times 10^{-5}$).
These small contributions of unstable states are consistent with
the estimation of the time of MI development.

\subsection{Accelerated deep optical lattice}
\label{accel_deep}

\subsubsection{LZ tunneling starting from a stable state.}

\begin{figure}[ht]
\epsfig{file=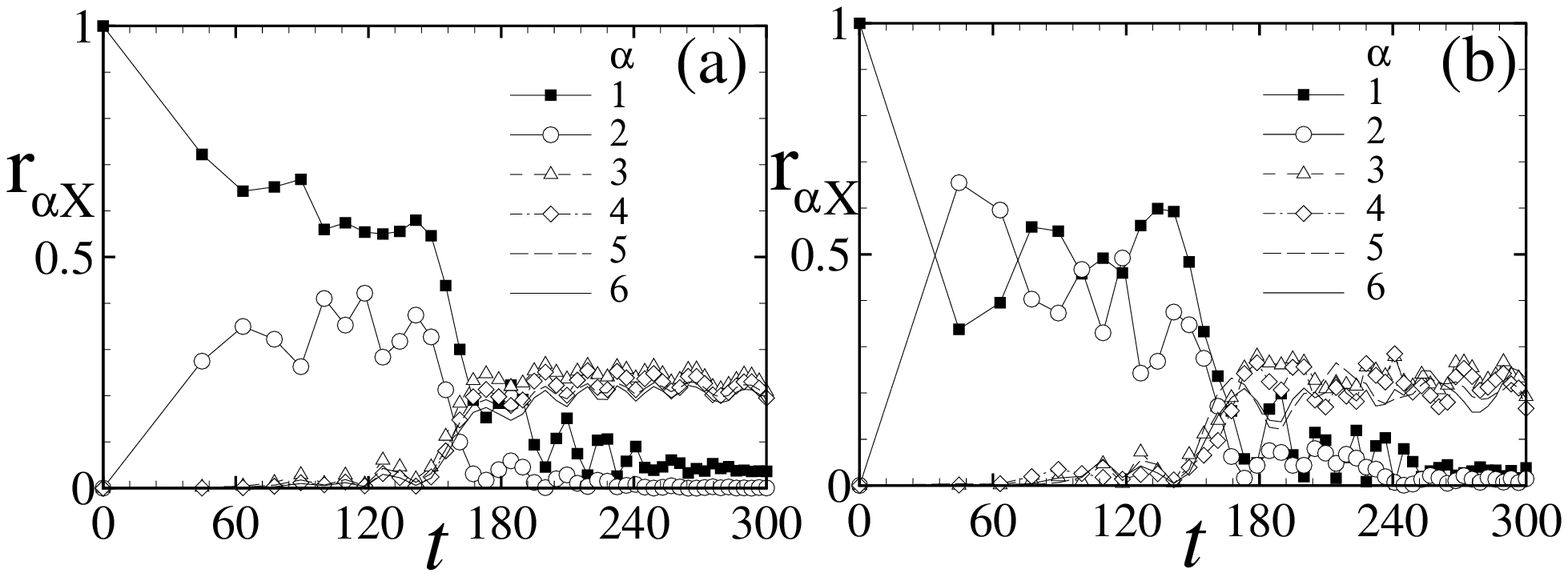,width=8cm,angle=0}
\epsfig{file=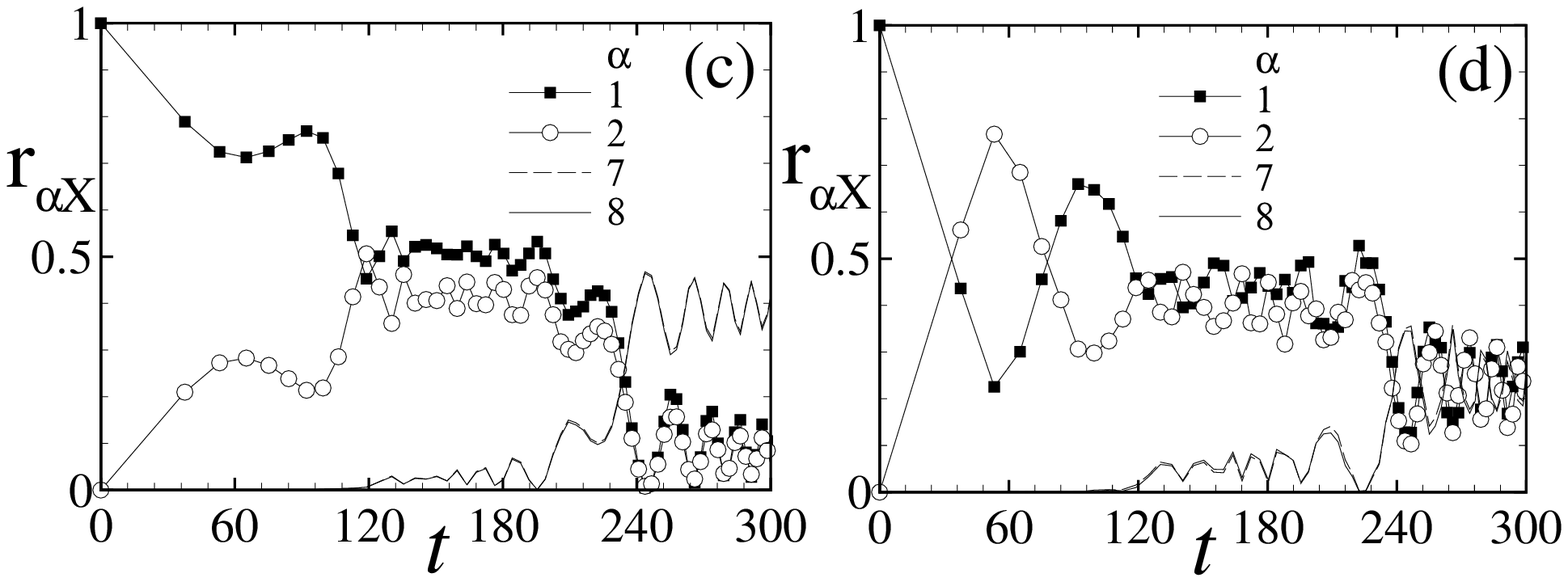,width=8cm,angle=0} \caption{Dynamics of populations of the
lowest bands in X-points with applied acceleration $a_x=5\times 10^{-4}$, $a_y=0$
in [(a),(b)] and  with $a_x=a_y=\frac5{\sqrt{2}}\times 10^{-4}$ in [(c),(d)]. In
[(a),(c)] the linear, $\sigma=0$, and in [(b),(d)] the nonlinear, $\sigma=1$,
regimes with $N=20$ are shown. Different symbols stand for populations of X-points
(the lines are guides for an eye).}
\label{fig_accel}
\end{figure}

Now we turn to the numerical study of tunneling in an accelerating OL. Along this
Section we  consider acceleration of a deep OL having the band structure
depicted in Fig.\ref{fig0}~(d). We start with the case when initially only a state
X$_1$ is populated and the acceleration is applied in the $x$-direction, ${\bf a}=(a_x,0)$.
The results are summarized in Fig.\ref{fig_accel}~(a),~(b) where
the populations are obtained at instants when the running OL having passed an integer number of period returns
to its initial position. In all our simulations we observed that the high symmetrical $\Gamma$-
and M-states do not become significantly populated during the dynamics.

Considering acceleration applied in the $x$-direction in the linear case
($\sigma=0$) in Eq.~(\ref{GPEa}) [see Fig.~\ref{fig_accel}~(a)], we observe three
different regimes. First,  during  the initial stage  the atoms tunnel into the
second band implying increase of the population of the X$_2$-state. This regime, at
$t\approx 60$ is changed to balanced distribution of atoms between X$_1$ and X$_2$
states, which lasts until $t\approx 150$. This stage of the evolution can be
explained due to the large energy difference between the states X$_2$ and
X$_{3,4,5,6}$. In our numerics the evolution is ended up in the particle transfer
to the upper states. We inspected the eight lowest bands and found that the process
is accompanied by developing rather complex spatial structures of the atomic
distribution. Since the states X$_{3,4,5,6}$ have very close energies, their
populations reveal almost synchronous behavior, as this is illustrated in
Fig.~\ref{fig_accel}~(a).

The nonlinear LZ tunneling has two additional features [see
Fig.~\ref{fig_accel}~(b)], compared to the linear one. The initial stage of the
tunneling from X$_1$ to X$_2$ is ended at earlier times, $t\approx 40$, and during
the second stage the dynamics reveals significant Rabi oscillations between the
states. Such oscillations occur due to  the cross-phase-modulation terms [see  the
two-state models (\ref{deep_latt_reson_inter}) or (\ref{deep_latt_nonreson_inter})]
and are characterized by a characteristic period $t_{Rabi}\approx 100$. Taking into
account that now the coupling coefficient $\kappa$ (\ref{kappa}) computed between
the two X-states $\kappa\approx 0.011$ we find that the analytical estimate for the
same time following from the two-state approximation yields $1/\kappa\approx 90$,
which is sufficiently good estimate, taking into account that, strictly speaking,
the two-state approximation is not valid any more. In the third stage one observes
tunneling of the atoms to the upper bands.

In order to understand how tunneling is affected by  the direction of acceleration,
we show in Fig.\ref{fig_accel}~(c),~(d) the results for $a_x=a_y$. We obtained them
in such a way that the total modulus of the acceleration is the same as in
Fig.\ref{fig_accel}~(a),(b). One can observe the two following changes in the
tunneling process. First, exchange between the first and second bands is much
stronger: during the second stage of evolution, $120\lesssim t\lesssim 200$, the
second band acquires larger population than the first band. Second, after the
instability is developed at $t>200$ the atoms start to populate  7-th and 8-th
bands while 3 -- 6-th bands remain negligibly populated.

This last phenomenon can be explained by the symmetry of the system. Indeed, let us consider the lowest eight BZs shown in Fig.~\ref{fig0}~(b). Departing from the X$_1$ point one finds the closed high-symmetry points located along the direction $(1,1)$, which coincides with the direction of acceleration, to be the X$_{7,8}$ points, while along the direction $(1,0)$, i.e. along the direction of acceleration shown in Fig.~\ref{fig_accel}~(a),(b), one finds the closed points X$_{3,4,5,6}$.

\subsubsection{LZ tunneling starting from  an unstable point.}

\begin{figure}[ht]
\epsfig{file=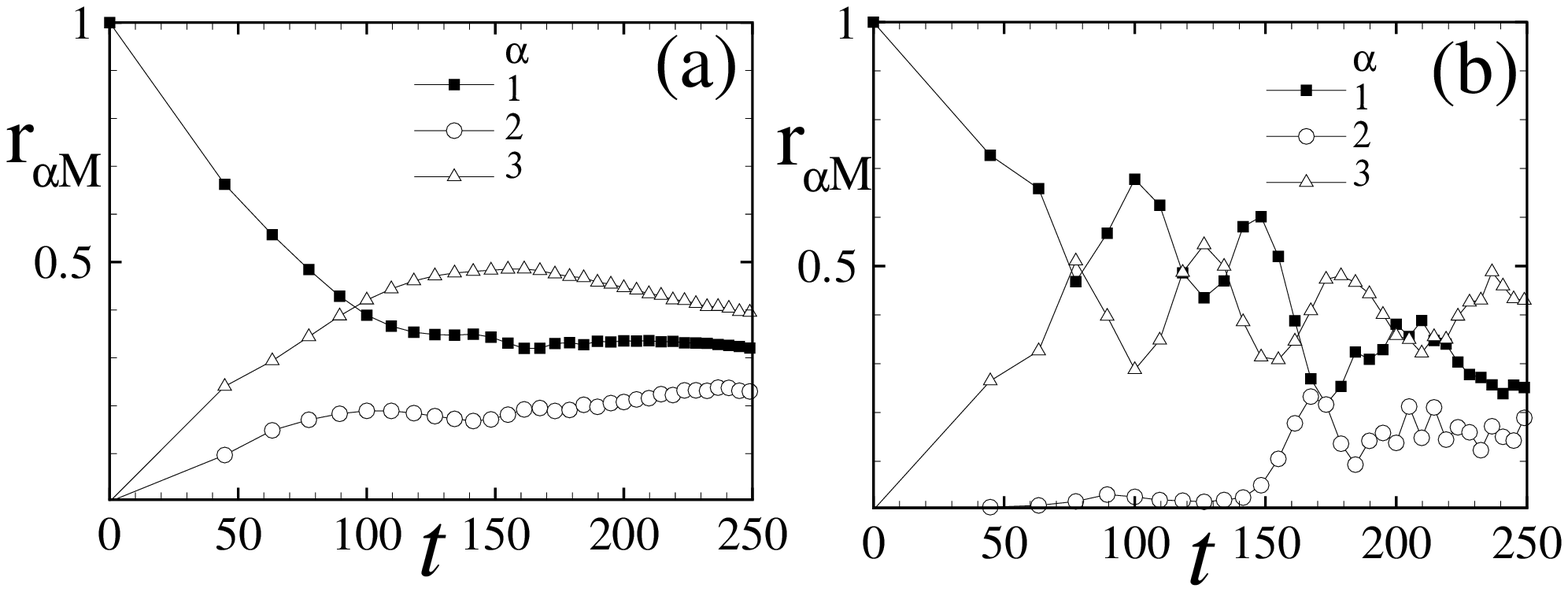,width=8cm,angle=0}
\epsfig{file=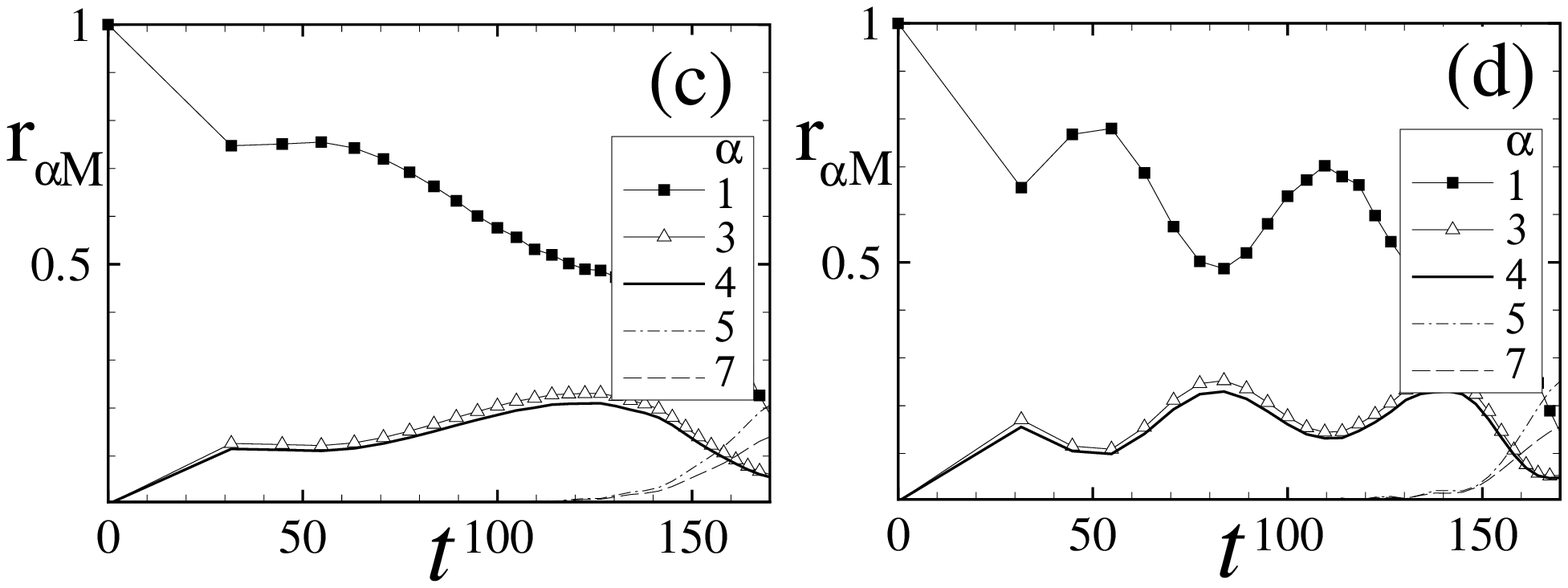,width=8cm,angle=0}
\caption{Dynamics of populations of the lowest bands in the M-point subject to the acceleration
$(a_x,a_y)=(5\times 10^{-4},0)$ with  $N=20$ in [(a),(b)] and to $a_x=a_y=5\times 10^{-4}$ with  $N=10$ in [(c),(d)].
In [(a),(c)] linear $\sigma=0$ and in [(b),(d)] nonlinear $\sigma=1$ regimes are shown. }
\label{fig_M1_ax_nl}
\end{figure}

Let us consider now the situation when initially all particles are placed in the unstable point M$_1$.
In Fig.\ref{fig_M1_ax_nl} we
present linear and nonlinear dynamics of the populations of the lowest bands  when the
acceleration is applied in (1,0) and in (1,1) directions.
In the panels (a) and (b) one observes that the three lowest bands participate in the exchange of particles,
which can be explained by the fact that the energy differences between the states M$_1$, M$_2$ and M$_2$, M$_3$
are of the  same order [see Fig.~\ref{fig0}~(d)]. Moreover, the population of the third band becomes larger than the
population of the two lowest band.

In the linear case the third stage of the evolution has not been reached for the
time $t\lesssim 250$ (this time was sufficient for transfer to upper bands in all
other cases considered so far). Thus the effect of the nonlinearity is much more
pronounced, than in the case of tunneling from the stable X-point. Now the
nonlinearity gives rise to the Rabi oscillations between the first and the third
bands accompanied by strong  attenuation of the population of the second band. One
can interpret this phenomenon by means of the definition of the nonlinear
coefficients $\chi_{\alpha\beta}$ [see (\ref{chi})], which are responsible for the
exchange of atoms between $\alpha$-th and $\beta$-th bands, and from the parity of
the BFs bordering a gap: BFs of the bands with different parity have also different
parity and, therefore, atomic transition induced by the nonlinearity between first
and third bands is more favorable than the oscillations between first and second
bands.

The results shown in Fig.~\ref{fig_M1_ax_nl}~(c),(d) indicate that change of the direction of the acceleration results
in significant variations in dynamics of populations.
Namely, as one can see in Fig.\ref{fig_M1_ax_nl}~(c), where the direction of acceleration coincides with the
direction of the symmetry axis (1,1), in the linear case there occurs a monotonic transfer of atoms from the first
band to the upper bands, populating almost equally M points in the third and forth bands, with tunneling to the upper
bands at later times. Remarkably, the nonlinearity does not affect much the tunneling in this case
[c.f. Figs.~\ref{fig_M1_ax_nl}~(b) and \ref{fig_M1_ax_nl}~(d)],  giving rise only to relatively weak
oscillations of the populations.

\subsubsection{LZ tunneling with broken symmetry between X and M states.}

\begin{figure}[ht]
\epsfig{file=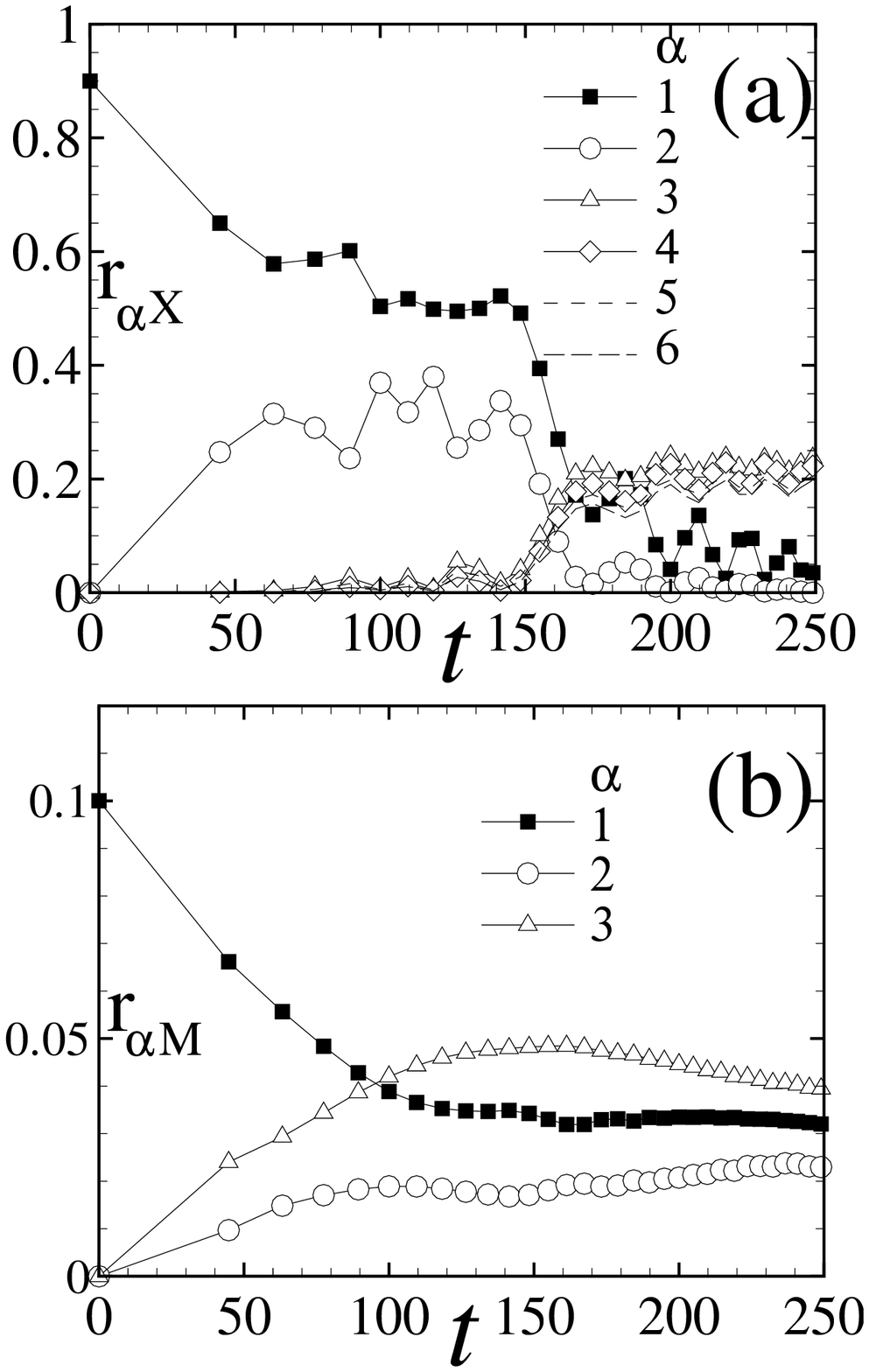,width=4cm,angle=0}\epsfig{file=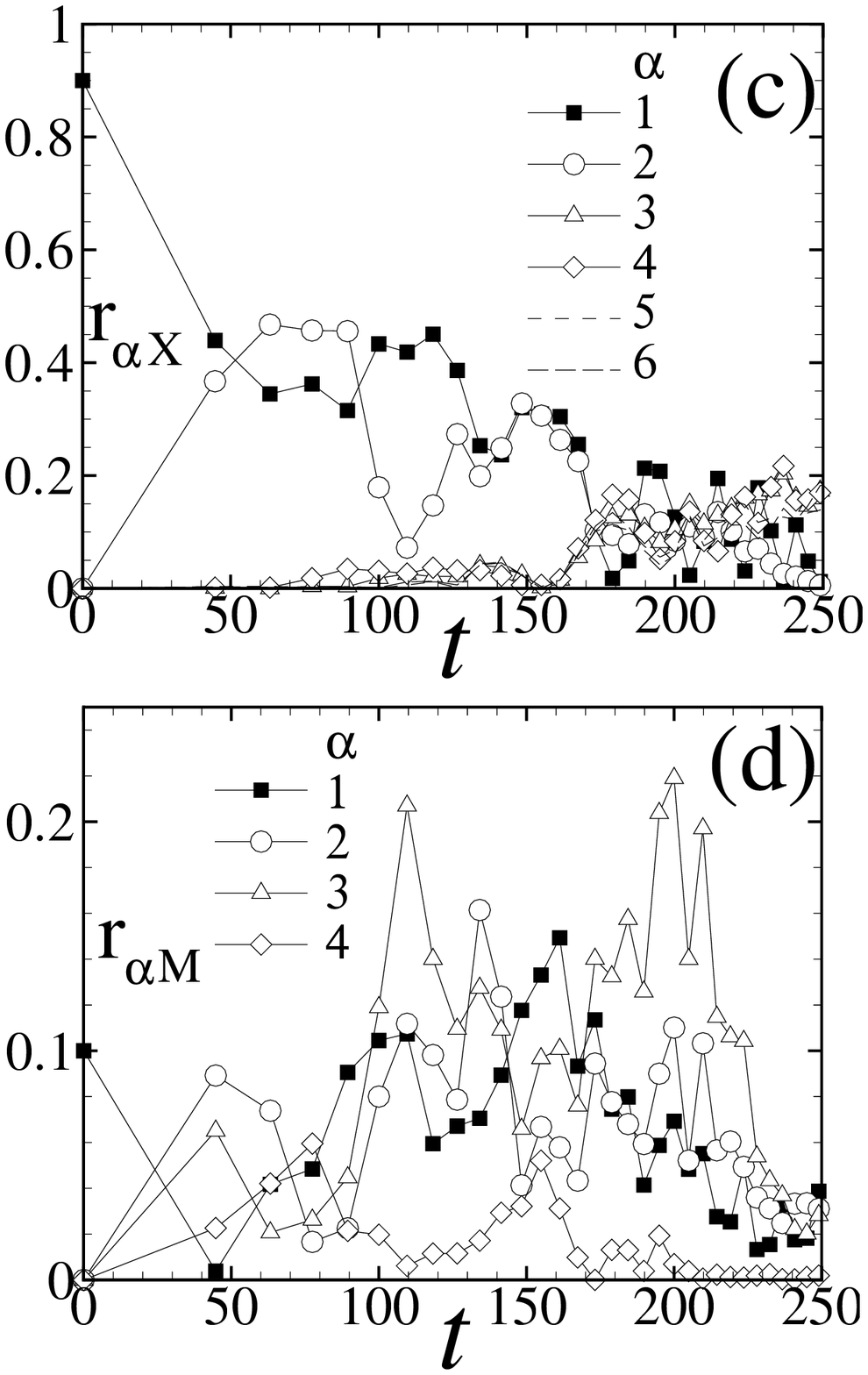,width=4cm,angle=0}
\epsfig{file=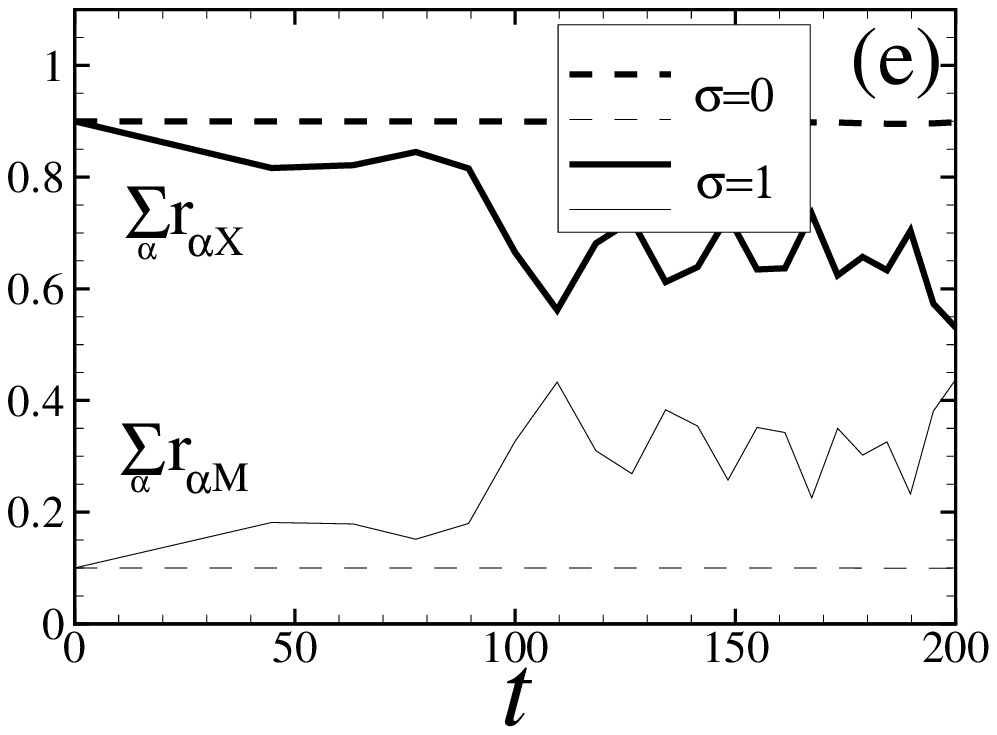,width=5cm,angle=0}
\caption{In  (a), (b)  linear, $\sigma=0$, and in  (c), (d) nonlinear, $\sigma=1$,  dynamics of the populations
of the lowest bands in X- and M-points with acceleration $(a_x,a_y)=(5\times 10^{-4},0)$ with $N=20$ are presented.
In the panels (a) and (c) populations in the X-point and in the panels (b) and (d) populations in the M-point
are shown. Initial distribution of the particles is r$_{1{\rm X}}=0.9$ and r$_{1{\rm M}}=0.1$.
Dynamics of the sum of the populations in the X- (thick solid and dashed lines) and in the M-points (thin solid
and dashed lines) in the linear (dashed lines) and nonlinear (solid lines) regimes is shown in (e).}
\label{fig_XM_ax_nl0_20}
\end{figure}

\begin{figure}[h]
\epsfig{file=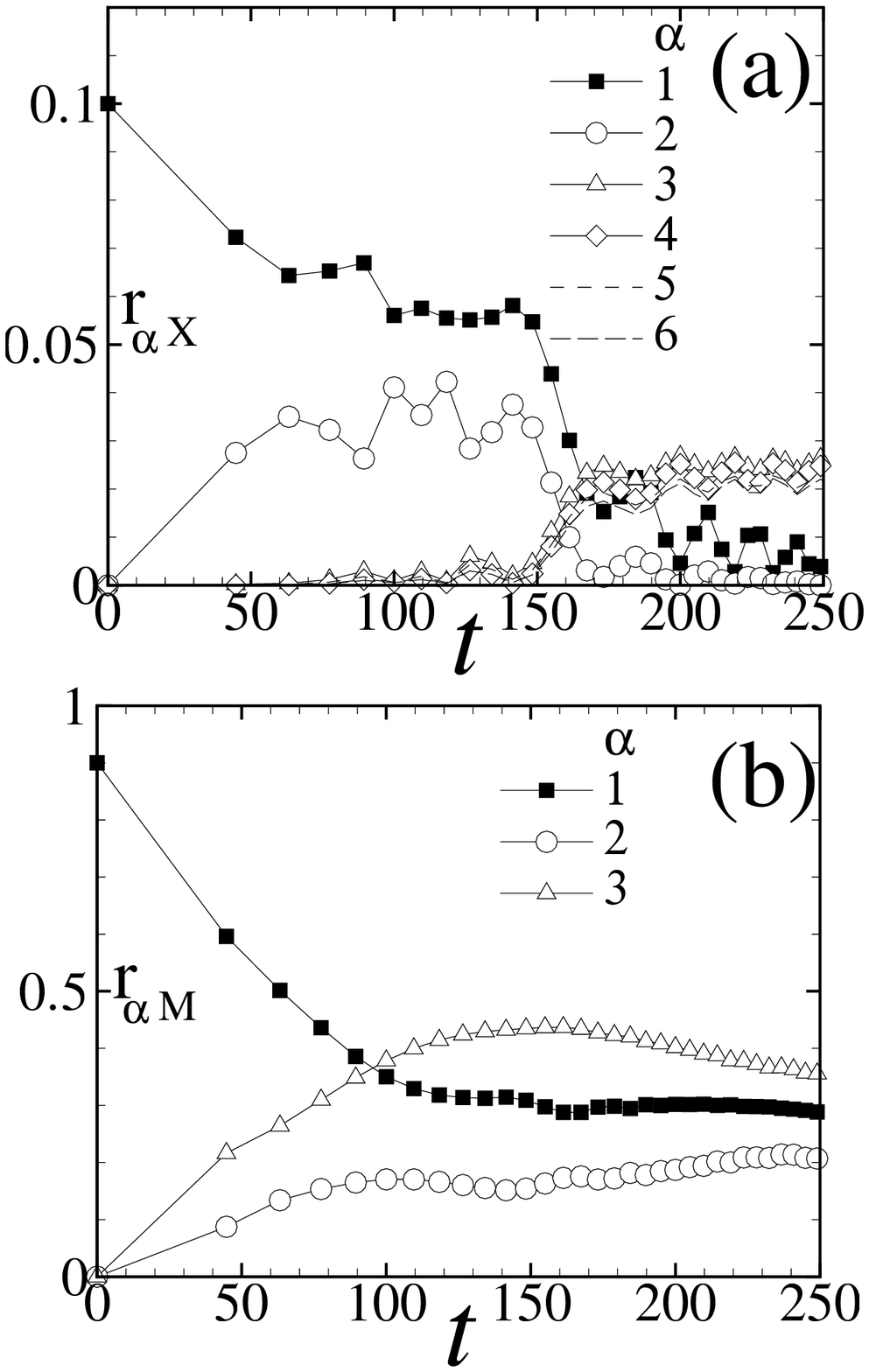,width=4cm,angle=0}\epsfig{file=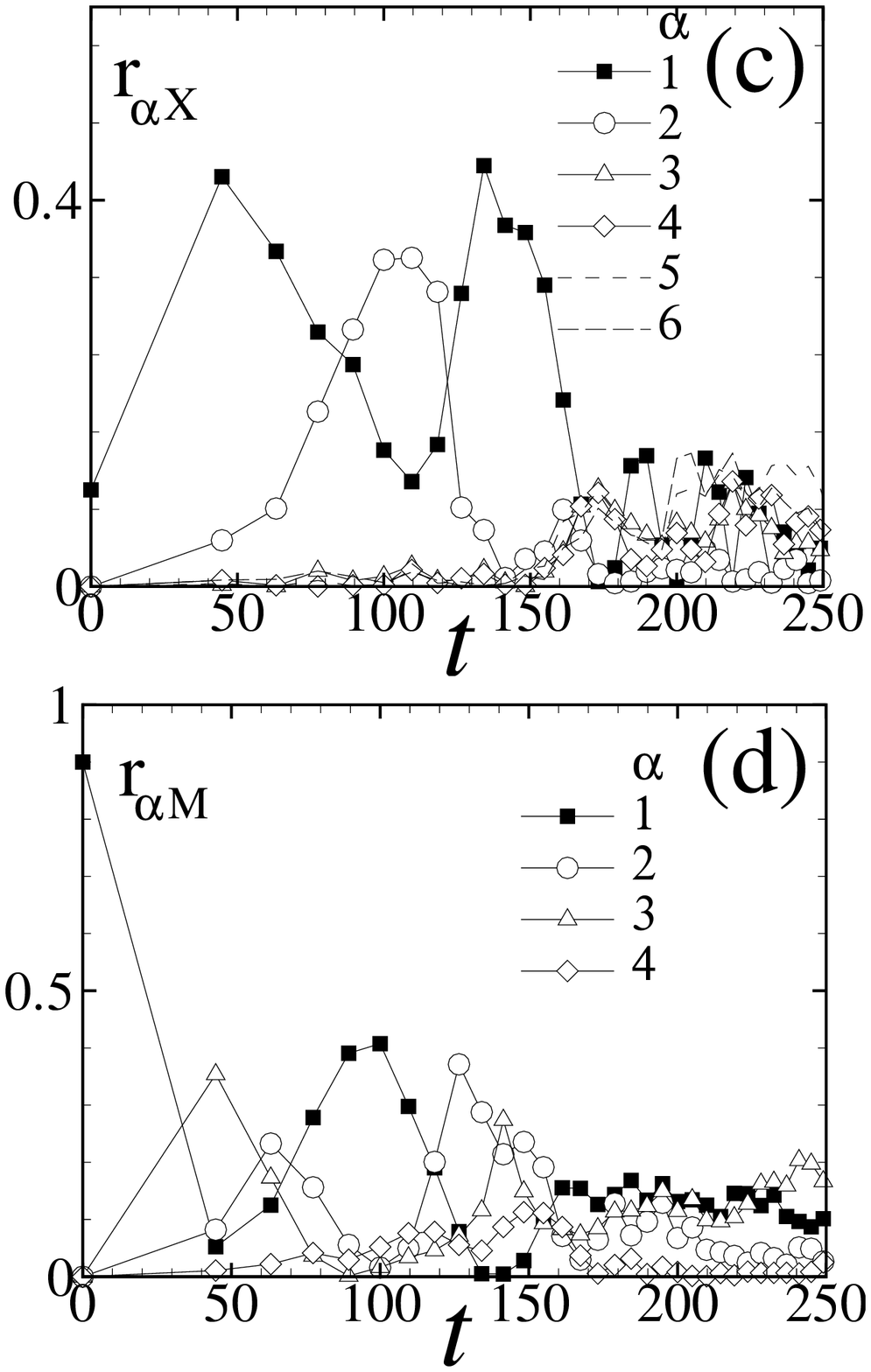,width=4cm,angle=0}
\epsfig{file=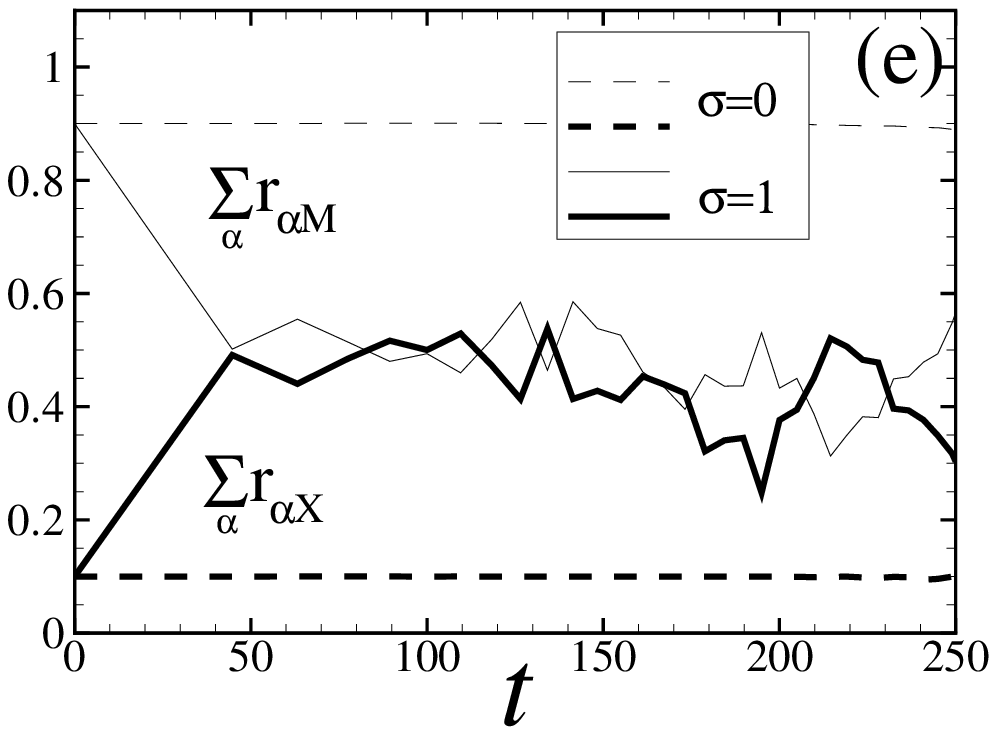,width=5cm,angle=0}
\caption{The same as in Fig.\ref{fig_XM_ax_nl0_20} with initial distribution of the particles r$_{1{\rm M}}=0.9$ and r$_{1{\rm X}}=0.1$.}
\label{fig_MX_ax_nl0_20}
\end{figure}

Now we focus on the situation when initially two states with different symmetry
occupied, namely, X- and M-states.  Specifically, in the case of the resonant
tunneling presented in Fig.\ref{fig0}~(d), we consider the two following
configurations: i) almost all particles are in the X$_1$-state while only small
number of particles are in the state M$_1$ [see Fig.~\ref{fig_XM_ax_nl0_20}] and
ii) almost all particles populate the state M$_1$  while only small number of
particles are in the X$_1$-state [see Fig.~\ref{fig_MX_ax_nl0_20}].

In order to verify whether there exists tunneling  between X- and M-states in we
calculated the sum of the populations of the eight lowest bands, $\sum_\alpha
r_{\rm X,M}$, in each symmetrical point in the linear and nonlinear regimes. The
results are presented in Figs.~\ref{fig_XM_ax_nl0_20}~(e) and
~\ref{fig_MX_ax_nl0_20}~(e). The thick dashed and thin dashed lines which represent
populations in X- and M-points indicate that particles in X- and M-points conserve
their initial population and thus behave independently. It means that in the case
of broken symmetry the  pure acceleration couples different bands but does not
couple  different symmetry points. In the presence of the nonlinearity (thick solid
and thin solid lines) the evolution of populations reveals Rabi oscillations of
populations between X- and M-points which occur due to cross-phase-modulation. We
note that this effect without acceleration was observed in \cite{BKK}.

The tunneling from the M- to the X-state, shown in  Fig.\ref{fig_MX_ax_nl0_20} is
significantly larger than in the case when particles tunnel from X- to M-state
presented in Fig.\ref{fig_XM_ax_nl0_20} what reveals extremely strong effect of the
instability which at very early times results in equal redistribution of atoms
between the two high-symmetry points.

%\end{widetext}

%\begin{widetext}
%\end{widetext}

%\input{NumericsShallow_26_02_07_rev}
\subsection{LZ tunneling in a shallow lattice}
\label{sec:numshallow}

\begin{figure}[ht]
\epsfig{file=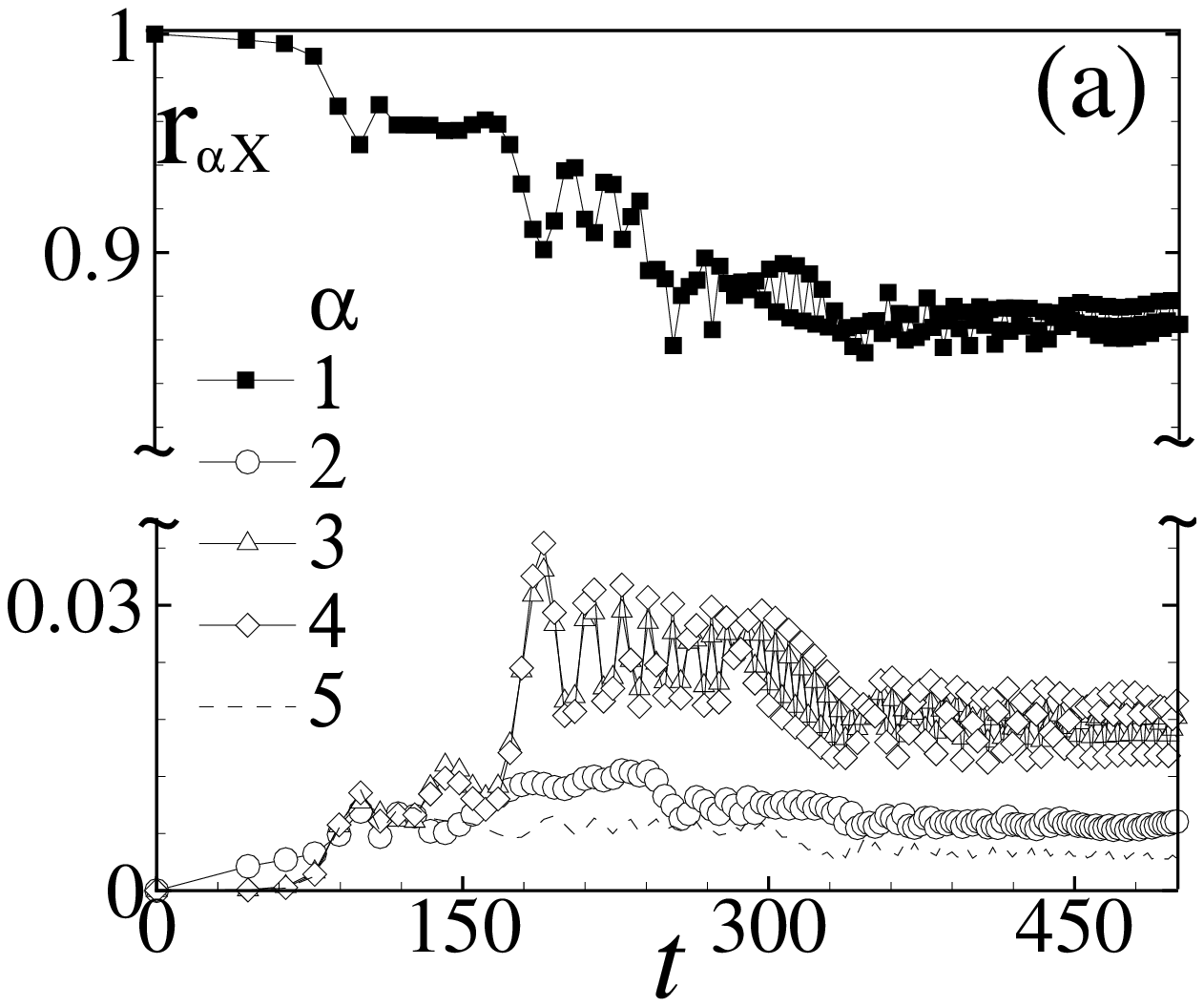,width=4cm,angle=0}
\epsfig{file=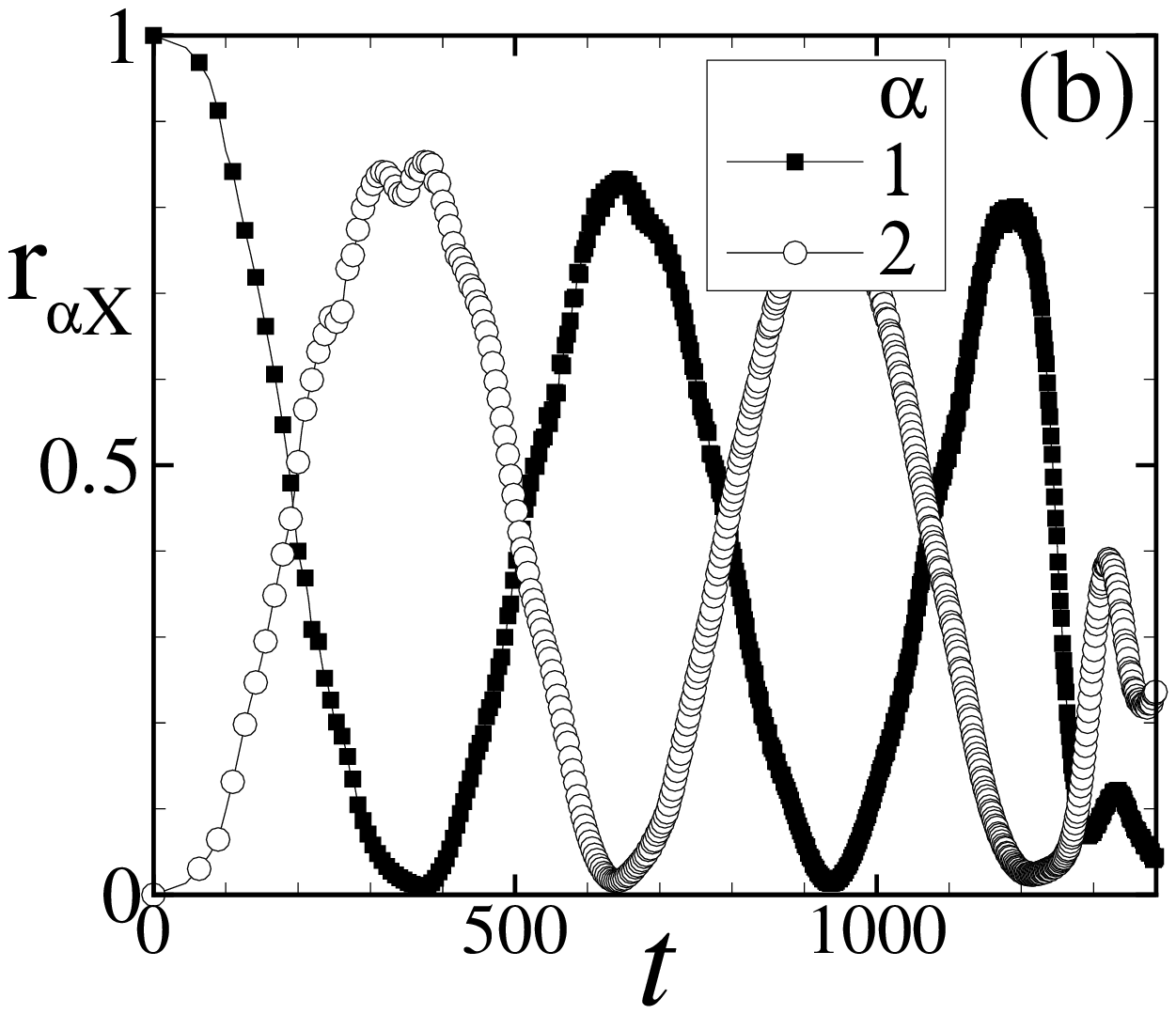,width=4cm,angle=0}
\caption{Dynamics of projections at the X-point
with applied acceleration $a_x=5\times 10^{-4}$ in (a) the linear case, $\sigma=0$
and in (b) the same as in (a) with the presence of nonlinearity, $\sigma=1$. Here $N=10$.
Other parameters are $V_0=-0.05$ and $\varepsilon=0.25$ [the band structure is shown in Fig.\ref{fig0}~(f)].
Different symbols denote the projections to the $\alpha$-bands
at the X-point calculated at time when the OL returns to the initial
position (thin lines are guides for an eye). }
\label{fig_accel_shallow}
\end{figure}

Now we turn to the case of the shallow OL corresponding to Fig.\ref{fig0}~(f).
We start with situation when all particles populate the X$_1$-point. By accelerating
the OL in the linear regime one can see in Fig.\ref{fig_accel_shallow}~(a),~(b)
that small number of particles succeeded to pass to the upper bands and
after some time interval (in our case $t\approx 300$) the population of the
first five bands becomes stable. The peculiarity of this process stems from the fact
that after $t\approx 200$ the population of the third and fourth bands starts to exceed
the population of the second band. This effect can be explained by the difference
in the effective mass of these bands by taking into account band structure from
Fig.\ref{fig0}~(f) the first, third and fourth bands have the negative effective
masses while the second and fifth bands have the positive ones.

One observes completely different dynamics in the nonlinear case with the same initial
conditions as in the linear one. In Fig.\ref{fig_accel_shallow}~(c) the population
of the particles in the presence of nonlinearity starts to oscillate between the first
and the second bands and only very small number of particles transfers into the upper
bands. After several periods of oscillations this dynamics evolves into the MI.

The dynamics in the shallow lattice case can be interpreted also in terms of the
Fourier powers instead of the Bloch band populations.  Indeed, as the theory
outlined in the Sec.~\ref{sec:shallow} predicts for the times smaller than the time
of the instability development, the X-point corresponds to the two-fold quasi 1D
resonance, hence predicting only two significant peaks in the Fourier space.
Namely, if we assume that the initial condition is broad enough in the real space,
e.g. a Gaussian possessing width covering many lattice periods, so that the Fourier
transform will be a narrow peak, then one can use the LZ model (\ref{EQ3}) (we
notice, that while Gaussian distribution of the density can be substituted by any
other one of the same spectral width, it can be created experimentally by first
loading the condensate  in a parabolic trap at low density, or alternatively
sufficiently small wavelength, and subsequently switching on the optical lattice
with simultaneous switching off the parabolic trap).

\begin{figure}[ht]
\epsfig{file=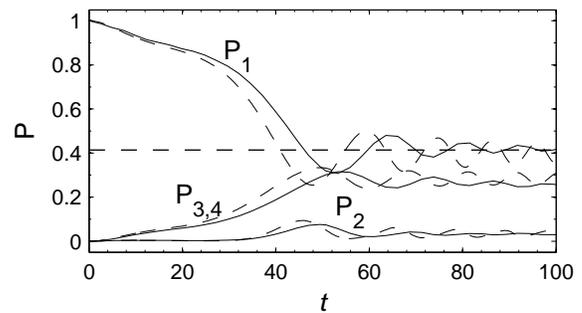}
\caption{Comparison of the LZ model with full PDE in the
linear case. We use the integrated powers of the Fourier transform of the order
parameter $\Psi$ (solid lines) and the squared absolute values of the LZ
coefficients (dashed lines). The horizontal line indicates the theoretical prediction
according to formula (\ref{EQ13}). Here $V_0 = 0.025$, $\vare = -1$ and $(a_x, a_y)= (0.01, 0)$.
The Gaussian initial condition for $\Psi$ with width of eight lattice
periods are used.}
\label{Zener1}
\end{figure}

We observe excellent agreement in the linear tunneling regime and a
good quantitative correspondence between the LZ models and the full PDE. For
instance, in Fig.~\ref{Zener1} the numerical results and theory for tunneling
at the M-point in the linear case are compared in the case of four-fold resonance.
One should keep in mind that for the initial value of the Bloch index we take into
account a time shift in the LZ model. In the nonlinear case, however, the theory
poorly corresponds to the numerical results, though there is excellent qualitative
correspondence easily observed in the Fourier space (for instance, in the number of
significant Fourier peaks).

\section{Conclusion}

In the present work we have carried out extensive analytical and numerical analysis of the nonlinear tunneling of Bose-Einstein condensates in  static and accelerating two-dimensional lattices. We focused on nonseparable potentials
possessing  different band gap structures: having a full gap, no gap, and satisfying
the resonant conditions with the full gap vanishing. We investigated different initial atomic
distributions in the highly symmetric states, which in the nonlinear case can be either
stable or modulationaly unstable, showing substantially different evolutions. We also
studied the effect of the direction of the lattice acceleration on the nonlinear tunneling.
Within analytical description of the process, we developed several two- and three-mode models, assuming that only a few states are occupied by the atoms.

We found that two-mode models accurately describe the tunneling before instabilities are developed, in the cases when either two respective states are initially populated or tunneling occurs between the two stable states, as this is the case of intra-band tunneling. The early stage evolution is dominated by the Rabi oscillations. Otherwise the two-mode models, although giving useful qualitative hints for understanding of tunneling and even estimates the characteristic times of the process, in general do not describe adequately all the peculiarities of the phenomenon. This deficiency can be explained by the fact that we deal with  spatially distributed systems, which in the nonlinear case develop instabilities
induced by the two-body interactions, which in their turn result in emergence of spatial coherent structures which are associated with population of higher bands.
Moreover, a two-dimensional band structure possesses such peculiarities as states having tensor of the inverse
reciprocal masses with components of different signs, what naturally diversifies dynamical scenarios of the tunneling and the symmetry of the developed patterns.
We also found that in accelerating lattices the instability phenomenon is much more pronounced and is developed at relatively early stages of evolution.  In the latter case we have shown that the direction of the lattice acceleration is a relevant physical parameter
significantly modifying atomic patterns emerging as a result of interplay between tunneling and instabilities.

\acknowledgments

VVK is grateful to D. E. Pelinovsky for letting know about the work \cite{21} prior its publication.
Authors thank M. Weitz for providing experimental results prior their publication.
VVK is pleased to thank the Department of Mathematics of the University of Castilla - La Mancha
for warm hospitality. V.A.B. was supported by the FCT grant SFRH/BPD/5632/2001. VVK acknowledges
support of Secretaria de Stado de Universidades e Investigaci\'on (Spain) under the grant SAB2005-0195.
 V.S.S. was supported by the research grants from CNPq and CAPES of Brazil. VAB and VVK were
supported by the FCT and FEDER under the grant POCI/FIS/56237/2004. Collaborative
work  was supported  by the agreement GRICES/Czech Academy of Sciences and by COST
P11 Action.

%\input{Appendix_02_04_07_rev}
%%%%%%%%%%%%%%%%%%%%%%%%%%%%%%%%%%%%%%%%%%%%%%%%%%%%%%%%%%%%%%%%%%%%%%%%%
%%%%%%%%%%%%%%%%%%%%%%%%%%%%%% APPENDICES %%%%%%%%%%%%%%%%%%%%%%%%%%%%%%%
%%%%%%%%%%%%%%%%%%%%%%%%%%%%%%%%%%%%%%%%%%%%%%%%%%%%%%%%%%%%%%%%%%%%%%%%%
\appendix

\section{Details of multiple-scale expansion}
\label{ap:multiscale}

\subsection{Some useful relations}

In this paper we use the relations as follows
\begin{eqnarray}
\label{I_fin_1}
   && \langle\varphi_{\alpha_1\bq_1}|\br|\varphi_{\alpha_2\bq_2}\rangle=\langle
    u_{\alpha_1\bq_1}|\br| u_{\alpha_2\bq_2}\rangle\delta_{\bq_1-\bq_2, \bQ}\,,
 \\
 \label{I_fin_2}
    &&\langle u_{1}|\nabla_{\bf q}|u_{2}\rangle =-i \langle \varphi_{1}|\br|
    \varphi_{2}\rangle
    =   -4i\frac{\langle \varphi_{1}|\nabla|\varphi_{2}\rangle}{E_2-E_1}
    \nonumber \\
    &&\qquad\qquad=   -4i\frac{\langle u_{1}|\nabla|u_{2}\rangle}{E_2-E_1}\,,
\\
\label{multiscale3}
       &&\langle   \varphi_{\alpha_i\bq_i}| \nabla^n|  \varphi_{\alpha_j\bq_j}\rangle
       = \delta_{\alpha_i\alpha_j}\delta_{\bq_i-\bq_j,\bQ}
       %\nonumber \\
       %\times
       \langle   \varphi_{\alpha_j\bq_j}| \nabla^n|  \varphi_{\alpha_j\bq_j}\rangle,
       \nonumber \\
\end{eqnarray}
(\ref{I_fin_2}) being valid for a periodic BFs bordering gap edges.

\subsection{Derivation of  Eqs. (\ref{deep_latt_reson_inter}).}
\label{ap:reson}

The higher order terms in the expansion (\ref{expan1}) are searched to be orthogonal to the leading order term $\psi_1$:   $ \langle\varphi_{1,2}\mid \psi_j\rangle=0$ ($j=2,3$).
 Therefore we represent $\psi_2$ as
\begin{eqnarray}
\label{psi2_2_ap}
    \psi_2=\sum_{\alpha,\bq}B_{\alpha\bq}(\br_1,t_1)\varphi_{\alpha\bq}e^{-iEt_0},
\end{eqnarray}
and choose the coefficients $B_{\alpha \bq}(\br_1,t_1)$ in the form
\begin{equation}
\label{multiscale4_1}
B_{\alpha \bq}
=B_\alpha^{(1)}
\delta_{\bq-\bq_1,\bQ}(1-\delta_{\alpha\alpha_1})
%\nonumber \\
+ B_\alpha^{(2)}
\delta_{\bq-\bq_2,\bQ}(1-\delta_{\alpha\alpha_2}),
\end{equation}
what justifies the formula (\ref{psi2_2}).

In order to obtain (\ref{A_1}) we substitute (\ref{psi2_2_ap}) with (\ref{multiscale4_1})
into (\ref{eigenvalue}) what yields
\begin{widetext}
\begin{eqnarray}
\label{muscale1_1}
 &&
\sum_{\alpha\neq\alpha_1}   \left[E_\alpha(\bq_1)-E\right] B_{\alpha}^{(1)}\varphi_{\alpha\bq_1}
%\nonumber \\
+\sum_{\alpha\neq\alpha_2}\left[E_\alpha(\bq_2)-E\right] B_{\alpha}^{(2)}\varphi_{\alpha\bq_2}
\nonumber \\
&&
=i\partial_{t_1}A_1\varphi_1 +i\partial_{t_1}A_2\varphi_2
%\nonumber \\
 +4\nabla_1A_1\cdot \nabla_0\varphi_1  +4\nabla_1A_2\cdot \nabla_0\varphi_2.
\end{eqnarray}
\end{widetext}
By applying $\langle\varphi_{1,2}\mid$ to the both sides of this equation and using
(\ref{multiscale3})  we obtain
\begin{eqnarray}
\label{multiscale5}
    \partial_{t_1}A_j-\bv\nabla_1A_j=0\quad \mbox{for}\quad j=1,2.
\end{eqnarray}
Next we apply  $\langle\varphi_{\alpha_j,\bq_{1,2}}|$ into (\ref{muscale1_1})  and use
(\ref{multiscale3}) and (\ref{A_1}). This yields for $j=1,2$ and $\alpha\neq\alpha_j$
\begin{eqnarray}
\label{bj}
    B_\alpha^{(j)}=4\frac{\langle u_{\alpha\bq_j}\mid\nabla_0\mid \varphi_j \rangle}
    {E_\alpha(\bq_j)-E}\nabla_1 A_j.
\end{eqnarray}

The equations of the third order (\ref{deep_latt_reson_inter}) are obtained
from (\ref{cL}) by using the orthogonality conditions
$\langle\varphi_{1,2}\mid \cF_3\rangle=0$, the explicit form of $\psi_2$
given by (\ref{psi2_2}) and (\ref{bj}), and the definition (\ref{mass}).

\subsection{Derivation of Eqs. (\ref{deep_latt_nonreson_inter}).}
\label{ap:sub:nonreson}

 Now the coefficients $B_{\alpha \bq}^{(j)}(\br_1,t_1) $   are as follows
\begin{eqnarray}
\label{multiscale4}
B_{\alpha \bq}^{(j)}
=
\delta_{\bq-\bq_j,\bQ}(1-\delta_{\alpha\alpha_j})B_\alpha^{(j)}, \quad j=1,2
\end{eqnarray}
what justifies the formula (\ref{psi2_1_ap}).

In order to obtain (\ref{A_1}) we substitute (\ref{psi2_1_ap}) with (\ref{multiscale4})
into (\ref{eigenvalue}), separate terms corresponding to the main harmonics
[$\propto\exp(-iE_{1,2}t_0)$] what yields ($j=1,2$)
\begin{eqnarray}
\label{muscale1}
% &&
\sum_{\alpha\neq\alpha_j}   \left[E_j-E_\alpha(\bq_j)\right]
B_{\alpha}^{(j)} \varphi_{\alpha\bq_j}=-i\partial_{t_1}A_j\varphi_j
\nonumber \\
-4\nabla_1A_j\cdot \nabla_0\varphi_j.
\end{eqnarray}
By applying $\langle\varphi_{1,2}|$ to the both sides of this equation and using
(\ref{vel_disc}) we compute Eqs. (\ref{multiscale5}).
 Next we apply  $\langle\varphi_{\alpha_j,\bq_{1,2}}|$ to (\ref{muscale1})  and
use  (\ref{multiscale3}) and (\ref{A_1}) . This yields for $j=1,2$ and $\alpha\neq\alpha_j$
\begin{eqnarray}
\label{bj_2}
    B_\alpha^{(j)}=4\frac{\langle u_{\alpha\bq_j}\mid\nabla_0\mid \varphi_j \rangle}
    {E_\alpha(\bq_j)-E_j}\nabla_1 A_j\,.
\end{eqnarray}

The equations (\ref{deep_latt_nonreson_inter}) are obtained from
(\ref{cL}) after separating the two main harmonics, using the orthogonality conditions
$\langle\varphi_{1,2}\mid \cF_3\rangle=0$, a particular explicit form of $\psi_2$,
(\ref{bj_2}), and the definition (\ref{mass}).

\end{document}